\documentclass[pdflatex,sn-vancouver-num]{sn-jnl}

\usepackage{graphicx}%
\usepackage{multirow}%
\usepackage{amsmath,amssymb,amsfonts}%
\usepackage{amsthm}%
\usepackage{mathrsfs}%
\usepackage[title]{appendix}%
\usepackage{xcolor}%
\usepackage{textcomp}%
\usepackage{manyfoot}%
\usepackage{booktabs}%
\usepackage{algorithm}%
\usepackage{algorithmicx}%
\usepackage{algpseudocode}%
\usepackage{listings}%

\newtheorem{theorem}{Theorem}[section]
\newtheorem{corollary}{Corollary}[section]
\newtheorem{proposition}{Proposition}[section]
\newtheorem{lemma}{Lemma}[section]
\newtheorem{example}{Example}[section]
\newtheorem{remark}{Remark}[section]
\newtheorem{definition}{Definition}[section]

\newcommand{\F}{\mathbb{F}} 
\newcommand{\Z}{\mathbb{Z}}
\newcommand{\zero}{{\mathbf{0}}}

\newcommand{\cS}{{\cal S}}
\newcommand{\cM}{{\cal M}}

\newcommand{\cH}{{\cal H}}

\begin{document}

\title{Weight distributions of free linear simplex codes over finite chain rings and their Gray map images}


\author[1]{\fnm{Cristina} \sur{Fernández-Córdoba}}\email{cristina.fernandez@uab.cat}

\author*[1]{\fnm{Sergi} \sur{Sánchez-Aragón}}\email{sergi.sanchez.aragon@uab.cat}

\author[1]{\fnm{Mercè} \sur{Villanueva}}\email{merce.villanueva@uab.cat}

\affil[1]{ \orgname{Universitat Autònoma de Barcelona, Barcelona, Spain}}


\abstract{A linear code of length $n$ over a finite chain ring $R$ with residue field $\F_q$ is an $R$-submodule of $R^n$. An $R$-linear code is a code over $\F_q$ (not necessarily linear) which is the generalized Gray map image of a linear code over $R$. These codes can be seen as a generalization of the linear codes over $\Z_{p^s}$ with $p$ prime and $s \geq 1$. In this paper, we present the construction of free linear simplex codes over $R$ and their corresponding $R$-linear simplex codes of types $\alpha$ and $\beta$.
Moreover, we determine the fundamental parameters of these codes, including their minimum Hamming and homogeneous distances, as well as their Hamming and homogeneous weight distributions. From the homogeneous weight distributions, we also derive the Hamming weight distributions of their corresponding Gray map images. We also study whether these simplex codes are optimal with respect to the Griesmer bound for codes over finite chain rings.}




\maketitle

\section{Introduction}

Let $R$ be a finite commutative ring with identity $1\not =0$. We say that $R$ is a {\it chain ring} if its ideals form a chain ordered by inclusion. Equivalently, $R$ is a chain ring if and only if it has a unique maximal ideal that is principal, that is, it is generated by one element, say $\gamma$. In fact, every ideal is of the form $\langle\gamma^{i}\rangle$ for $i\in \{0,\dots, s\}$, where $s$ is the minimum natural number such that $\gamma^{s}=0$, which is called the {\it nilpotency index}. 

 As $R$ is finite, its {\it residue field} $R/\langle \gamma \rangle$ is a finite field, so $R/\langle \gamma \rangle\cong \mathbb{F}_{q}$ for some prime power $q=p^{r}$. From this point on, the residue field of $R$ is denoted as $\F_q$.
 There is a canonical projection homomorphism from $R$ onto $\F_q$. Denote by $\bar{r}$ the image of an element $r\in R$ under this projection. Let $T=\{e_{0},\dots,e_{q-1}\}\subseteq R$ such that $\overline{e_{i}}\neq\overline{e_{j}}$ for all $i,j \in \{0,1,\dots,q-1\}$, $i\neq j$. 
The set $T$ is a set of coset representatives of $\F_q$, so $\F_q=\{\overline{e_{0}},\dots,\overline{e_{q-1}}\}$.  For convenience, we choose $e_{0}=0$ and $e_1=1$. For every $r\in R$, there are unique $r_{0},\dots, r_{s-1}\in T$ such that $r=\sum_{i=0}^{s-1}r_{i}{\gamma^{i}}$ \cite{SalageanChain}.        
We refer to $[r_{0},\dots,r_{s-1}]_{\gamma}$ as the {\it $\gamma$-adic representation} of $r$.

A {\it code} over $R$ of length $n$ is a nonempty subset $C_R$ of $R^{n}$, and it is a {\it linear code} over $R$ if it is an $R$-submodule of $R^{n}$.
Given a code $C_R$ over a ring $R$ of length $n$  and a distance function $d_{w}$ defined by a weight function $w$, we define the {\it minimum distance} of $C_R$ as $d_{w}(C_R)=\min\{d_{w}(x,y)\mid x,y\in C_R, \, x\not =y \}$. We also define the {\it minimum weight} of $C_R$ as $w(C_R)=\min\{ w(x)\mid x\in C_R, \, x\not =(0,\dots,0) \}$. If $W(C_R)$ denotes the maximum weight of the codewords in $C_R$ and $A_{i}$ denotes the number of codewords with weight equal to $i$, for $i\in \{0,\dots,W(C_R)\}$, we say that the values $A_{0},\dots,A_{W(C_R)}$ are the {\it weight distribution} of $C_R$ for the weight $w$. We define the {\it distance distribution} of $C_R$ for a distance $d_{w}$ analogously. It is well-known that, when $C_R$ is linear, the weight and distance distributions for $w$ and $d_{w}$, respectively, coincide. 

We are interested in two weights widely used in coding theory, namely, the Hamming weight and the homogeneous weight. The first one, the {\it Hamming weight}, is $w_{H}(x)=|\{i\in\{1,2,\dots,n\}\mid x_{i}\neq 0\}|$, where $x=(x_{1},\dots,x_{n}) \in R^n$ and $R$ is a ring. Then, the {\it Hamming distance} is defined as $d_H(x,y)=w_H(x-y)$ for any $x,y\in R^n$.
The Hamming weight distribution of a code $C_R$  of length $n$ is often condensed in its {\it weight enumerator}, which is the polynomial defined as $W_{C_R}(X,Y)=\sum_{i=0}^{W(C_R)}{A_{i}X^{n-i}Y^{i}}$.
The second one, the {\it homogeneous weight}, is defined for any finite chain ring $R$ as follows. Given $x\in R$,
$$w_{Hom}(x)=\begin{cases} 0 & \text{if $x=0$,} \\
  (q-1)q^{s-2} & \text{if $x\neq0$ and $x\notin\langle \gamma^{s-1}\rangle$,}\\
  q^{s-1} &\text{if $x\neq0$ and $x\in\langle \gamma^{s-1}\rangle$.}
\end{cases}$$
It is easily extended to $R^{n}$ component-wise as  $w_{Hom}(x)=\sum_{i=1}^{n}{w_{Hom}(x_{i})}$, where $x=(x_{1},\dots, x_{n}) \in R^n$. Then, the {\it homogeneous distance} is defined as $d_{Hom}(x,y)=w_{Hom}(x-y)$ for any $x,y\in R^n$.

As shown in \cite{SalageanChain}, any linear code $C_R$ over a finite chain ring $R$ of length $n$ is equivalent, via permutation of columns and multiplication of columns by units, to a linear code with a generator matrix of the form
\begin{equation} \label{eq:salagean matrix}
\left(\begin{array}{cccccc}
I_{t_{1}} & A_{1,1} & A_{1,2} &\cdots & A_{1,s-1} & A_{1,s}\\
0  & \gamma I_{t_{2}} & \gamma A_{2,2} &\cdots &\gamma A_{2,s-1} & \gamma A_{2,s}\\
\vdots & \vdots &\vdots &\ddots &\vdots &\vdots\\
0 & 0 & 0 &\cdots & \gamma^{s-1}I_{t_{s}} &\gamma^{s-1}A_{s,s}
\end{array}\right),
\end{equation}
where $A_{i,j}$ is a matrix with entries over $R$ for $1\leq i\leq s$, $i\leq j\leq s$, and $I_{t_i}$ is the $t_i\times t_i$ identity matrix for $1\leq i\leq s$.
Moreover, since the values $t_{1},\dots,t_{s}$ are unique \cite{SalageanChain}, we say that the linear code $C_R$ is of \textit{type} $(n;t_{1},\dots,t_{s})$. A linear code $C_R$ of type $(n;t_{1},\dots,t_{s})$ has size $|C_R|=q^{\sum_{i=1}^{s}{(s-(i-1))t_{i}}}.$
It is worth noting that the type offers a characterization of free codes, since  a linear code $C_R$ over $R$ is free if and only if its type is $(n;t_{1},0,\dots,0)$.

Let $H$ be a matrix over $\F_{q}$. We say that $H$ is a {\it generalized Hadamard (GH) matrix}, $H=H(q,\lambda)=(h_{ij})$, of order $n=q\lambda$ over $\F_{q}$ if $H$ is a $q\lambda\times q\lambda$ matrix such that, for every $i, j$,  $1\leq i<j\leq q\lambda$,
each of the multisets $\{h_{is}-h_{js} \mid 1\leq s \leq q\lambda\}$ contains every element of $\F_{q}$ exactly $\lambda$ times. A code $C_q$ over $\F_{q}$ is called a {\it generalized Hadamard (GH) code} if $C_q=\bigcup_{a\in\F_{q}}{(F+(a,\dots,a))}$, where $F\subseteq \F_q^{q\lambda}$ is the code consisting of the rows of a GH matrix $H(q, \lambda)$, and $F+(a,\dots,a)=\{x+(a,\dots,a)\mid x\in F\}$. Such a code has length $q\lambda$ and $q^{2}\lambda$ codewords.
A linear code $C_R$ over $R$ such that its Gray map image $\Phi_{R}(C_R)$ is a GH code is called an {\it $R$-additive GH code}. 
The recursive construction of a family of these codes given in \cite{Dipak23} is of special interest, as the families of codes we want to study are closely related to it.
This notion is of special interest for the case when $R=\Z_{p^{s}}$, as simplex codes over these rings and their Gray map images are closely related to GH codes \cite{Dipak23,Hadamard2}. 

As shown in \cite{Hisometries}, the success of the Gray map defined in \cite{Z4} being an isometry from $\Z_{4}$ with the Lee distance to $\Z_{2}^{2}$ with the Hamming distance cannot, in general, be replicated. That is, there is no isometry $f:\Z_{p^{s}}\longrightarrow \Z_{p}^{s}$ for the metric space $(\Z_{p}^{s}, d_{H})$ with $s>2$. However, other isometries can be built, such as 
the generalized Gray maps from $\Z_{p^{s}}$ to $\Z_{p}^{p^{s-1}}$ given in \cite{Dipak22,Dipak23,Carlet2k,Hadamard2,HadamardZ8,Nonlinearity}. 
More generally, in \cite{Greferath}, a Gray map for an arbitrary finite chain ring $R$ was defined as a function $\Phi_{R}:R\longrightarrow  \mathbb{F}_{q}^{q^{s-1}}$ given by
\vspace{-0.15truecm}
\begin{equation} \label{eq:Greferath}
\Phi_{R}(r)=(\overline{r_{0}},\dots,\overline{r_{s-1}})M_{s-1}^{R},
\end{equation}
where  $[r_{0},\dots, r_{s-1}]_\gamma$ is the $\gamma$-adic representation of $r$, and $M_{s-1}^{R}=\left(\begin{array}
{c} 
Y_{s-1}\\
1\dots1  
\end{array}\right)$ is an $s\times q^{s-1}$ matrix, where $Y_{s-1}$ is a matrix whose columns are all the elements of $\F_q^{s-1}$ and the last row consists entirely of ones.
Note that $M_{s-1}^{R}$ is a generator matrix of the first-order Reed-Muller code, $RM(s-1,1)$, over $\mathbb{F}_{q}$. In fact, $\Phi_R(R)$ is also a linear GH code of length $q^{s-1}$ over $\F_q$, which comes from a generalized Hadamard matrix $H(q,q^{s-2})$, known as the Sylvester Hadamard matrix \cite{MacWilliamsSloane}.
We can extend $\Phi_{R}$ to a Gray map on $R^{n}$, denoted also by $\Phi_R$, applying it component-wise. 

Let $C_R$ be a linear code over $R$ of type $(n;t_{1},\dots,t_{s})$. Then, $C_q=\Phi_R(C_R)$ is a code over $\F_q$ of length $N=n\cdot q^{s-1}$ with $|C_q|=|C_R|$ codewords. We also say that $C_q$ is of type $(n;t_{1},\dots,t_{s})$. Note that $C_q$ is not necessarily linear over $\F_q$. A code $C_q$ over $\F_q$ is called an {\it $R$-linear code} if $C_q=\Phi_R(C_R)$ for some linear code $C_R$ over $R$. 
Studying the preimage $C_R$ of an $R$-linear code $C_q=\Phi_R(C_R)$ is of interest, as, for example, the homogeneous weights of the codewords in $C_R$ are the Hamming weights of the codewords in $C_q$. This is due to the fact that $\Phi_{R}:(R^n,d_{Hom})\longrightarrow (\F_{q}^{nq^{s-1}},d_{H})$ is an isometry \cite{Greferath}.

The so-called $q$-ary simplex codes are  $[\frac{q^k-1}{q-1},k,q^{k-1}]$ linear codes, defined as the duals of the well-known Hamming codes over $\F_q$ \cite{MacWilliamsSloane}.
Since the publication of \cite{Z4} in 1994, the study of codes over rings has become increasingly important. As a result, several versions of simplex codes over rings have been defined and studied, mainly over finite chain rings. 

The families of $\Z_{2^{s}}$-additive simplex codes of types $\alpha$ and $\beta$  were introduced in \cite{Gupta}, where their Hamming and homogeneous weight distributions were studied.
In \cite{Nonlinearity}, the relationship between $\Z_{2^s}$-additive simplex codes and $\Z_{2^{s}}$-additive Hadamard codes was established, and the linearity of their Gray map images was investigated, showing that these images are not linear in general. These $\Z_{2^{s}}$-additive Hadamard codes, and more generally the $\Z_{p^{s}}$-additive GH codes, were constructed and studied in \cite{Dipak22,Dipak23,Hadamard2,HadamardZ8,AdrianZp}. They are linear codes over $\Z_{p^s}$ whose corresponding $\Z_{p^s}$-linear codes are GH codes, which are not necessarily linear. 

The main motivation for this paper is to construct and study two families of linear simplex codes,  $\cS_k^{\alpha}$ and $\cS_k^{\beta}$, over any finite chain ring $R$, generalizing the constructions of $\Z_{2^s}$-additive simplex codes of types $\alpha$ and $\beta$ introduced in \cite{Gupta}. Specifically, $\cS_k^{\alpha}$ is defined by a generator matrix whose columns consist of all elements of $R^k$, whereas $\cS_k^{\beta}$ is obtained from $\cS_k^{\alpha}$ by puncturing certain positions to eliminate proportional columns. When $R$ is a finite field $\F_q$, $\cS_k^{\alpha}$ is the subcode of the first-order Reed-Muller code over $\F_q$ obtained by removing the row of its generator matrix whose entries are all equal to $1$, while $\cS_k^{\beta}$ coincides with the classical $q$-ary simplex code.

In the literature, several other families of simplex or simplex-related codes have been defined over different rings. We show that these families do not coincide with the family introduced in this paper. A family of linear codes over $\Z_{m}$, called $\Z_m$-simplex codes, was introduced in \cite{Durairajan2015,ChellaPandian2015}. For the particular case $m=p^s$, the Hamming weight distribution of these codes was given in \cite{Durairajan2015} for dimension $2$, while  their covering radius with respect to the Hamming metric was determined in \cite{ChellaPandian2015}. More recently, the covering radii of these simplex codes over $\Z_{p^s}$
 with respect to the homogeneous and Lee metrics were investigated in \cite{SoleCovRad}. These codes have length $(m^k-1)/(m-1)$, as in the case of classical $q$-ary simplex codes, and can be obtained by puncturing  $\cS_k^{\beta}$. Hence, the $\Z_m$-simplex codes and the codes  $\cS_k^{\beta}$ have different lengths. 
 
Simplex codes of types $\alpha$, $\beta$, and $\gamma$ over $\mathbb{Z}_6$ were introduced in \cite{Senarysimplex}, where their Hamming weight distributions were also determined. The generator matrices of the codes of types $\alpha$ and $\gamma$ are constructed in the same way as those of $\cS_k^{\alpha}$ and $\cS_k^{\beta}$, respectively. On the other hand, codes of type $\beta$ are constructed differently and are not related to either of these two families.

More recently,  unit simplex codes, unit simplex codes of type $\alpha$, and zero-divisor simplex codes over $\mathbb{Z}_m$ were introduced in \cite{unitsimplex}. The Hamming weight distributions of these codes were determined for dimension $3$ when $m=p^{s}$. The unit simplex codes are obtained by puncturing the simplex codes constructed in \cite{Durairajan2015}, while the unit simplex codes of type $\alpha$ are obtained by puncturing the unit simplex codes. Therefore, both families can be obtained by puncturing $\cS_k^{\beta}$. Moreover, the zero-divisor simplex codes over $\mathbb{Z}_{p^{s}}$ can also be obtained  by puncturing $\mathcal{S}_{k}^{\beta}$. Consequently, these three families have shorter lengths than the codes $\cS_k^{\alpha}$ and  $\cS_k^{\beta}$. 

Simplex codes of type $\alpha$ were introduced over the finite chain ring $\mathcal{R}=\F_2[X]/(X^{3})$, and their Hamming and Lee weight distributions were investigated in \cite{SimplexPolynomialrngs}. These codes  are constructed in the same way as $\cS_k^\alpha$ for this particular chain ring.  In the same work, a Gray map for codes over this ring was defined and the properties of the corresponding binary images were studied. This Gray map is linear, and therefore all resulting binary images are linear codes, unlike the Gray map images considered in this paper, which are not necessarily linear. 

Finally, simplex codes of types $\alpha$ and $\beta$ were introduced in \cite{SimplexRq} over the rings $\mathcal{R}_m=\F_2[u_1,\dots,u_m]/\langle u_i^{2}=0, u_i u_j=u_j u_i \rangle$ for $m\geq 1$, which are local rings but not chain rings. Lee and homogeneous weights are defined in this setting, and the corresponding weight distributions of the simplex codes are derived. These codes are constructed in a way analogous to $\cS_k^\alpha$ and $\cS_k^\beta$, respectively, but adapted to the rings $\mathcal{R}_m$.
Gray maps associated with each weight are also introduced, and their binary images are linear codes, unlike the Gray map images considered in this paper, which are not necessarily linear. 
Table 1 summarizes this survey of simplex-related codes.
{\scriptsize
\begin{table}[h]
    \renewcommand{\arraystretch}{1}
    \begin{tabular}{|l|c|c|l|}
         
        \hline
         Family &Ring &Length &Relation to $\cS_k^{\alpha}$ and $\cS_k^{\beta}$  \\
      \hline
         $\Z_{2^{s}}$-additive simplex codes &$\Z_{2^{s}}$ &$\alpha:2^{sk}$ ,   &Equal when $R=\Z_{2^{s}}$ \\
          of type $\alpha$ and $\beta$ \cite{Gupta} &&$\beta:2^{(s-1)(k-1)}(2^k-1)$& \\
        \hline
          $\Z_m$-simplex codes \cite{Durairajan2015,ChellaPandian2015} & $\Z_m$ &$\frac{m^k-1}{m-1}$ &Punctured from $\cS_k^{\beta}$, for $m=p^s$ \\
         \hline
        Senary simplex codes  &$\Z_6$ &$\alpha:6^{k}$ , &Same construction for $\alpha$ and $\gamma$,  \\  of type $\alpha,\beta$ and $\gamma$ \cite{Senarysimplex} &&$\beta:\frac{(2^k-1)(3^k-1)}{2}$, &$\beta$ differs\\ &&$\gamma:2^{k-1}(2^k+3^k)$& \\
          \hline
          Unit simplex codes \cite{unitsimplex} &$\Z_m$ &$\frac{(\varphi(m)+1)^k-1}{\varphi(m)}$ &Punctured from $\cS_k^{\beta}$, for $m=p^s$ \\
          \hline
         Unit simplex codes $\alpha$ \cite{unitsimplex} &$\Z_m$ &$\frac{\varphi(m)^k-1}{\varphi(m)-1}+\varphi(m)^{k-2}$ &Punctured from $\cS_k^{\beta}$, for $m=p^s$\\
          \hline
         Zero divisor simplex  &$\Z_m$ &$\frac{(m-\varphi(m))^k-1}{(m-\varphi(m))-1}$ &Punctured from $\cS_k^{\beta}$, for $m=p^s$ \\
         codes \cite{unitsimplex} &&& \\
         \hline
         Simplex codes \cite{SimplexPolynomialrngs} &$\mathcal{R}$  &$8^k$ &Equal to $\cS_k^\alpha$ when $R=\mathcal{R}$ \\
         \hline
         Simplex codes  &$\mathcal{R}_{m}$ &$\alpha:2^{2^m}$,   &Same construction for $\alpha$ and $\beta$ \\
         of type $\alpha$ and $\beta$ \cite{SimplexRq} &&$\beta:\frac{2^{(2^{2^m}-1)(k-1)}}{2^k-1}$& \\
        \hline 
         
    \end{tabular}
     \caption{Summary of simplex-related codes  and their relationship with $\cS_k^{\alpha}$ and $\cS_k^{\beta}$, where $\mathcal{R}=\F_2[X]/(X^{3})$, $\mathcal{R}_m=\F_2[u_1,\dots,u_m]/\langle u_i^{2}=0, u_i u_j=u_j u_i \rangle$, and $\varphi$ is Euler's totient function.}
\end{table}
}
\vspace{-0.75truecm}

In this paper, we study the fundamental parameters of $\cS_k^{\alpha}$ and $\cS_k^{\beta}$, as well as those of their Gray map images under $\Phi_{R}$. We also determine the Hamming and homogeneous weight distributions of these codes, and use these results to obtain the Hamming weight distributions of their Gray map images. Finally, we study the optimality of $\cS_k^{\alpha}$ and $\cS_k^{\beta}$ with respect to the Griesmer bound for codes over rings introduced in \cite{ShiromotoG}. These results were previously determined for $\Z_{2^{s}}$-additive simplex codes in \cite{Gupta}; here, we extend them to arbitrary finite chain rings. 
 The simplex codes $\cS_k^{\beta}$ over finite chain rings were previously defined in terms of certain multisets of points in projective Hjelmslev geometries in \cite{HonoldLandjev}, where their Hamming weight distribution was also determined. We show that both families coincide and we provide an alternative proof of this result using a different approach. Furthermore, we determine their homogeneous weight distribution, which was not considered in \cite{HonoldLandjev} and from which the Hamming weight distribution of their Gray images can be derived.

An interesting consequence of these results is  that the Gray map images of simplex $\alpha$ codes are one-weight, while those of simplex $\beta$ codes are two-weight. It is important to note that two-weight codes obtained as Gray map images of linear codes over certain chain rings have been used in cryptography, particularly in secret-sharing schemes \cite{JianfaQIAN2012,graymapimagessecretsharing}. In steganography \cite{ApplicationstegoZ4,RR11}, some of the proposed codes are nonlinear codes arising as Gray map images of codes over rings. The Gray map images of the $\mathbb{Z}_{2^{s}}$-additive simplex codes are nonlinear, except when $s=2$ and $k=1$ \cite{Nonlinearity}. We conjecture that, in general, the Gray map images of $\mathbb{Z}_{p^{s}}$-additive simplex codes are also nonlinear for any prime $p>2$.

The paper is organized as follows. In Section \ref{sec:construction}, we construct  the families of  free linear simplex codes of types $\alpha$ and $\beta$ over any finite chain ring $R$, denoted by $\cS_k^\alpha$ and $\cS_k^\beta$, respectively. In Section \ref{sec:distributionsimplex}, we determine the minimum Hamming and homogeneous distances of the codes in these families, together with their Hamming and homogeneous weight distributions. From the homogeneous weight distributions, we also derive  the Hamming weight distributions for their corresponding Gray map images, $\Phi_R(\mathcal{S}_k^\alpha)$ and $\Phi_R(\mathcal{S}_k^\beta)$. We also study whether these simplex codes are optimal with respect to the Griesmer bound introduced in \cite{ShiromotoG}. In Section \ref{sec:SimplexZps}, we focus on the particular case when $R=\Z_{p^s}$ with $s\geq 1$ and $p$ a prime number. We show that free $\Z_{p^{s}}$-additive simplex codes $\cS_k^\alpha$ of type $\alpha$ are closely related to free $\Z_{p^{s}}$-additive GH codes.
Finally, in Section \ref{sec:conclusion}, we present our conclusions and discuss directions for future research.

\section{Construction of linear simplex codes} \label{sec:construction}

In this section, we present the construction of two families of linear simplex codes over finite chain rings, giving a recursive construction for their generator matrices. They are a generalization of the ones introduced in \cite{Gupta} when $R=\Z_{2^s}$.

We now introduce some notation used in some proofs along the paper. Let $R$ be a ring. If a vector $x\in R^n$ appears as an entry of another vector $y$, it means that its coordinates are included as individual components of $y$. For example, if $x=(1,2,3)$, then $y=(x,4)=(1,2,3,4)$. Given a vector $x\in R^n$ and a positive integer $k$, the vector $(x,\dots,x)$, where $x$ is repeated $k$ times, is denoted by $x^{(k)}$. For example, $(1,2,3,1,2,3)=(1,2,3)^{(2)}$.  The vector $x=(a,\dots,a)\in R^n$ is also denoted by $\mathbf{a}^{(n)}$, or just  $\mathbf{a}$ if the length is clear from the context. For instance, $(1,1,1)=\mathbf{1}^{(3)}$  or just $\mathbf{1}$.

Let $R$ be a finite chain ring. Recall that $T=\{e_{0},\dots,e_{q-1}\}\subseteq R$ is a set of coset representatives of $\F_q$. We consider that the elements of $T$ are ordered so that $e_{0}=0$ and $e_1=1$. Given $x,y\in R$ such that $x=[x_{0},\dots,x_{s-1}]_{\gamma}$ and $y=[y_{0},\dots,y_{s-1}]_{\gamma}$, we say that $x\geq_{\gamma}y$ if and only if $x_{i}>y_{i}$ as elements in $T$, where $i$ is the highest index such that $x_{i}\neq y_{i}.$ From now on, we consider the elements of a finite chain ring $R$ to be $\{\rho_{0},\rho_{1},\dots, \rho_{q^{s}-1}\}$, which are listed in ascending order. From  the order definition, it follows that $\rho_{0}=0$ and $\rho_{1}=1.$


\begin{remark}\label{rmk:idealcount}
Recall that every ideal  of $R$ is of the form $\langle\gamma^{j}\rangle$. Note that $|\langle \gamma^{j} \rangle|=q^{s-j}$, for every $j\in\{0,\dots,s\}.$
In particular,  $|R|=q^{s}.$
\end{remark}

\begin{definition} \label{def:simplexalpha} 
Given a finite chain ring $R$, we define the matrix $G_{1}^{\alpha}=\left(\begin{array}{ccccc} \rho_{0} &\rho_{1} &\rho_{2} &\dots &\rho_{q^{s}-1}\end{array}\right).$ Then, for $k>1$, we define the matrix 

$$G_{k}^{\alpha}=\left(\begin{array}{ccccc}
\boldsymbol{\rho_{0}} &\boldsymbol{\rho_{1}} &\boldsymbol{\rho_{2}} & \dots & \boldsymbol{\rho_{q^{s}-1}}\\
G_{k-1}^{\alpha} & G_{k-1}^{\alpha} & G_{k-1}^{\alpha} & G_{k-1}^{\alpha} & G_{k-1}^{\alpha}
\end{array}\right).$$ 
The linear code over $R$ generated by $G^\alpha_k$, denoted by $\mathcal{S}_{k}^{\alpha}$, is called a linear simplex $\alpha$ code with $k$ generators.
\end{definition}

Another way to construct the code $\mathcal{S}_{k}^{\alpha}$, without using an iterative construction, is to take as its generator matrix the matrix whose columns are all the elements of $R^k$, ordered appropriately to obtain the same matrix  $G_{k}^{\alpha}$  as in Definition \ref{def:simplexalpha}. Since $G_{k}^{\alpha}$ contains all the elements of $R^{k}$ as columns, the code has length $|R^k|=|R|^k=q^{sk}$.

\begin{example}\label{ex:SimplexAlphaZ9} 
Let $R=\Z_9$. We have that
$G_{1}^{\alpha}=\left(\begin{array}{ccccccccc}
0&1&2&3&4&5&6&7&8
\end{array}\right),$
and
$$G_{2}^{\alpha}=\left(\begin{array}{ccccc}
000000000 &111111111 &222222222 &\cdots  &888888888 \\
012345678 &012345678  &012345678 &\cdots &012345678
\end{array}\right).$$
\end{example}

\begin{example}\label{ex:SimplexAlphaGR42}
Let $R=GR(4,2)\cong\Z_{4}[x]/(x^{2}+x+1)$. Consider the elements of $R$ in ascending order, that is, $R=\{0,1,\omega,\omega+1,2,3,2+\omega,3+\omega,2\omega,2\omega+1,3\omega,2\omega+2,2\omega+3,3\omega+2,3\omega+3\},$ where $\omega$ is a root of the basic irreducible polynomial $x^2+x+1$ over $\Z_{4}.$
Then, we have that
$G_{1}^{\alpha}=(0 \;\; 1 \;\; \omega \;\; \omega+1 \;\; 2 \;\; 3 \;\; 2+\omega\;\; 3+\omega  \ \dots \ 
3\omega+2 \;\; 3\omega+3),$
and 
$$G_{2}^{\alpha}=\left(\begin{array}{cccccc}
\mathbf{0}^{(16)} &\mathbf{1}^{(16)} &\boldsymbol{\omega}^{(16)} &\cdots &\boldsymbol{3\omega+3}^{(16)}\\
0\, 1\, \omega\,\dots\,3\omega+3 &0\,1\,\omega\,\dots\,3\omega+3 &0\,1\,\omega\,\dots\,3\omega+3 &\cdots &0\,1\,\omega\,\dots\,3\omega+3
\end{array}\right).$$
\end{example}

\begin{proposition} \label{prop:2.1}
Let $s_{i}^{k}$ denote the $i$th row of $G_{k}^{\alpha}$, $i\in\{1,2,\dots,k\}$. Then, 
$s_{i}^{k}=\big(\boldsymbol{\rho_{0}}^{(q^{s(k-i)})},\boldsymbol{\rho_{1}}^{(q^{s(k-i)})},\dots,\boldsymbol{\rho_{q^{s}-1}}^{(q^{s(k-i)})}\big)^{(q^{s(i-1)})}.$
\end{proposition}

\begin{proof}
For $k=1$, the result is trivial, since $G_{1}^{\alpha}$ has only one row that lists all elements in $R$. 
Now, consider $k\geq 2$. For $i=1$, by the construction of $G_{k}^{\alpha}$, we have $s_{1}^{k}=(\boldsymbol{\rho_{0}}^{(q^{s(k-1)})},\boldsymbol{\rho_{1}}^{(q^{s(k-1)})},\dots,\boldsymbol{\rho_{q^{s}-1}}^{(q^{s(k-1)})})$.  For $2\leq i\leq k$, by construction, we have $s_{i}^{k}=(s_{i-1}^{k-1})^{(q^{s})}.$ By applying this relation iteratively $i-1$ times, we obtain $s_{i}^{k}=(s_{1}^{k-(i-1)})^{(q^{s(i-1)})}.$ Therefore, we have $s_{i}^{k}=(\boldsymbol{\rho_{0}}^{(q^{s(k-i)})},\boldsymbol{\rho_{1}}^{(q^{s(k-i})},\dots,\boldsymbol{\rho_{q^{s}-1}}^{(q^{s(k-i)})})^{(q^{s(i-1))})}.$
\end{proof}

\medskip
Let $\langle \gamma \rangle=\{a_{0}\gamma,\dots,a_{q^{s-1}-1}\gamma\}$, where the elements are in ascending order. Note that $\{a_{0},\dots, a_{q^{s-1}-1}\}$ is a set of representatives of $R/\langle \gamma^{s-1} \rangle$ and $a_0=0$.

\begin{definition} \label{def:2.5}
Given a finite chain ring $R$, we  define the matrix $G_{1}^{\beta}=(1)$. Then, for $k>1$, we define the matrix 
\begin{equation}\label{eq:simplexbetam}
G_{k}^{\beta}=\left(\begin{array}{cccccc} 
\mathbf{1} &a_0 \boldsymbol{\gamma} &a_1\boldsymbol{\gamma} &a_{2}\boldsymbol{\gamma} & \dots &a_{q^{s-1}-1}\boldsymbol{\gamma}\\
G_{k-1}^{\alpha} & G_{k-1}^{\beta} & G_{k-1}^{\beta} & G_{k-1}^{\beta} &\dots &G_{k-1}^{\beta} 
\end{array}\right). 
\end{equation}
The linear code over $R$ generated by $G_{k}^{\beta}$, denoted by $\mathcal{S}_{k}^{\beta}$, is called a linear simplex $\beta$ code with $k$ generators.
\end{definition}

\begin{example}\label{ex:SimplexBetaZ9}
Let $R=\Z_9$. We have that
$G_{1}^{\beta}=\left(1\right)$ and
$$G_{2}^{\beta}=\left(\begin{array}{cccc}
1\,1\,1\,1\,1\,1\,1\,1\,1 &0 &3 &6\\
0\,1\,2\,3\,4\,5\,6\,7\,8 &1 &1 &1
\end{array}
\right).$$
\end{example}

\begin{example}\label{ex:SimplexBetaGR42}
Consider $R=GR(4,2)$ as in Example \ref{ex:SimplexAlphaGR42}, we have that
$G_{1}^{\beta}=(1),$ and
$$G_{2}^{\beta}=\left(\begin{array}{ccccc}
\mathbf{1}^{(16)} &0 &2 &2\omega &2\omega+2\\
0\,1\,\omega\,\dots\,3\omega+3 &1 &1 &1 &1
\end{array}\right).$$
\end{example}

Recall that the columns of the generator matrix $G_{k}^{\alpha}$ of  $\mathcal{S}_{k}^{\alpha}$ contain exactly all the elements of $R^{k}$.
The construction of $G_{k}^{\beta}$ is similar but removing some of the columns of $G_{k}^{\alpha}$, in order to assure that no two columns are multiples of each other, as we show in the next proposition.

\begin{proposition} \label{prop:simplexbetaprojective}
Let $G_k^{\beta}$ be the generator matrix of $\cS_k^{\beta}$ over $R$, with $k\geq1$, given in Definition \ref{def:2.5}. Let $\{g_{i}\}_{1\leq i\leq n}$ be the set of its columns. Then, we have that
\begin{description}
\item[(i)] For every $c\in R^k \setminus \{\zero\}$, there exist $j\in\{1,\ldots,n\}$ and $\lambda\in R\setminus\{0\}$ such that $c=\lambda g_j$.
\item[(ii)] If $g_i=\lambda g_j$ for some $i,j\in\{1,\ldots,n\}$ and $\lambda\in R$, then $\lambda=1$ and $i=j$.
\end{description}
\end{proposition}
\begin{proof} For both items, we use the recursive construction of $G_k^{\beta}$ given in Definition~\ref{def:2.5}.

(i) We prove the result by induction on $k\geq 1$. For $k=1$, the result is clear since $G_1^{\beta}=(1)$. Let $k>1$ and assume that the result holds for $k-1$. Consider $c=(c_1,c_2)\in R^{k}\setminus\{\zero\}$, where $c_1\in R$ and $c_2\in R^{k-1}$. If $c_2=\zero$, then $c=(c_1,0,\dots,0)$ with $c_1\not =0$. Since $g_1=(1,0,\dots,0)^{T}$ is the first column of $G_k^{\beta}$, $c=c_1g_1$ and the result follows with $\lambda =c_1$. Thus, we may assume that $c_2\neq \zero$. By the induction hypothesis, there exist $\mu\in R\setminus\{0\}$ and a column $g$ of $G_{k-1}^{\beta}$ such that $c_2=\mu g$. 
If $c_1=0$, then $c=\mu(0,g)$. Since $(0,g)^T$ is a column of $G_k^\beta$, there exists $j\in\{1,\ldots,n\}$ such that $c=\mu g_j$ and the result follows with $\lambda=\mu$. 
Hence, we may further assume that $c_1\neq 0$. Then, we have $c=(c_1,\mu g)$ where both $\mu$ and $c_1$ are nonzero. Since 
$R$ is a finite chain ring, either $\mu\mid c_1$ or $c_1\mid\mu$. We therefore distinguish two cases.

If $\mu\mid c_1$, then $c_1=\mu a$ for some $a\neq 0$. Consequently, $c=\mu(a,g)$. If $a\in \langle \gamma \rangle$, then $(a,g)^{T}$ is a column of $G_k^{\beta}$. Hence, there exists  $j\in\{1,\ldots,n\}$ such that $c=\mu g_j$, and the result follows with $\lambda=\mu$. Otherwise, $a$ is a unit, since $R$ is a finite chain ring. In this case, $(a,g)=a(1,a^{-1}g)$. By construction of $G_k^{\beta}$, and since the columns of $G_{k-1}^{\alpha}$ include all the elements of $R^{k-1}$, $(1,a^{-1}g)^{T}$ is a column of $G_k^{\beta}$. Hence, there exists $j\in\{1,\ldots,n\}$ such that $c=\mu a(1,a^{-1}g)=\mu a g_j$ and the result holds with $\lambda=\mu a$.
Finally, if $c_1\mid \mu$, then $\mu= c_1 b$ for some $b\neq 0$.  As before, since $(1,bg)^{T}$ is a column of $G_{k}^{\beta}$, 
there exists $j\in\{1,\ldots,n\}$ such that $c=(c_1,c_1b g)=c_1g_j$  and the result follows with $\lambda=c_1$.

(ii) Again, we prove the result by induction on $k\geq 1$. For $k=1$, it is clear, since $G_{1}^{\beta}=(1)$. Let $k>1$, and assume that the result is true for $k-1$.  By reduction to the absurd, suppose there are $i\neq j$, and $\lambda\in R\setminus\{0,1\}$ such that $g_{i}=\lambda g_{j}$. We consider four different cases. First, if $i,j>q^{s(k-1)}$, as the last $k-1$ coordinates of $g_{i}$ and $g_{j}$ are columns of $G_{k-1}^{\beta}$, then equation $g_{i}=\lambda g_{j}$ contradicts the induction hypothesis.
Second, if $i,j\leq q^{s(k-1)}$, as the first coordinate of $g_{i}$ and $g_{j}$ is equal to one, then we obtain that $1=\lambda$, so $g_i=g_j$ which is a contradiction.
Third, if $i\leq q^{s(k-1)}$ and $j>q^{s(k-1)}$, looking at the first coordinate of $g_{i}$ and $g_{j}$, then we have $1=\lambda z$, where $z$ is an element of $\langle \gamma \rangle$. However, this also means that $z$ is a unit, which is a contradiction. Finally, we consider the case when $i>q^{s(k-1)}$ and $j\leq q^{s(k-1)}$. Since $g_{i}=\lambda g_{j}$, looking at the first coordinate, we have that $z=\lambda$, where $z \in \langle \gamma \rangle$. Thus, all the entries in $\lambda g_{j}$ are also in $\langle \gamma \rangle$. By induction, one can prove that any column of $G_{k}^{\beta}$ has at least one entry equal to one. This means that there is at least one entry in $g_{i}$ equal to one, which is a contradiction.  
Therefore, either $i=j$ or $\lambda=1$.
\end{proof}

\begin{proposition} \label{prop:3.2}
For $k\geq1$, the linear simplex $\beta$ code $\mathcal{S}_{k}^{\beta}$ has length 
\vspace{-0.25truecm}
$$L_{\beta}(k)=q^{(s-1)(k-1)}\frac{q^{k}-1}{q-1}.$$
\end{proposition}
\begin{proof}
We prove the result by induction over $k\geq 1$.
For $k=1$, it is true since the length is $L_{\beta}(1)=1$.  If $k>1$, by construction, we have that 
\begin{equation} \label{eq:recsimp}
L_{\beta}(k)=q^{s(k-1)}+q^{s-1}L_{\beta}(k-1).
\end{equation}
Indeed, the first term in (\ref{eq:recsimp})  corresponds to the number of columns of $G_{k-1}^{\alpha}$, as follows from its construction in Definition \ref{def:simplexalpha}. The second term corresponds to the number of elements of  $\langle \gamma \rangle$, which is $q^{s-1}$ by Remark \ref{rmk:idealcount}.
Assume $L_{\beta}(k-1)=q^{(s-1)(k-2)}\frac{q^{k-1}-1}{q-1}$. Now, we have that
$L_{\beta}(k)=q^{s(k-1)}+q^{s-1}L_{\beta}(k-1)
            =q^{s(k-1)}+q^{s-1}q^{(s-1)(k-2)}[(q^{k-1}-1)/(q-1)]=
            q^{k-1}q^{(s-1)(k-1)}+q^{(s-1)(k-1)}[(q^{k-1}-1)/(q-1)]
 =q^{(s-1)(k-1)}[((q-1)q^{k-1}+q^{k-1}-1)/(q-1)]
 =q^{(s-1)(k-1)}[(q^{k}-1)/(q-1)].$
\end{proof}

From Propositions \ref{prop:simplexbetaprojective} and \ref{prop:3.2}, the columns of $G_{k}^{\beta}$ can be seen as points in  a projective space over $R$. They correspond to the elements of a projective Hjelmslev space over $R$ of dimension $k-1$ \cite{HonoldLandjev}. This fact implies that $\mathcal{S}_{k}^{\beta}$ is permutation equivalent to the code $\operatorname{Sim}(k,R)$ defined in \cite{HonoldLandjev} using a geometrical approach.  Moreover, it was shown that the so-called fat linear codes are related to the hyperplanes of Hjelmslev spaces. A \textit{fat} code $C_R$ over $R$ of length $n$ is defined as a code such that, for every position $i\in\{1,\dots, n\}$, there is a codeword $c=(c_1,\dots,c_n)$ such that $c_{i}$ is a unit. 
By Proposition \ref{prop:simplexbetaprojective}, the linear simplex $\beta$ code $\cS_{k}^{\beta}$ over $R$ is a fat code. Therefore, the results established in \cite{HonoldLandjev} apply to $\cS_{k}^{\beta}$ and allow for a geometric derivation of its Hamming weight distribution. However, in this work, all results concerning $\cS_{k}^{\beta}$ are proved using an independent approach. The results for $\cS_{k}^{\alpha}$ are not affected, as it is not a fat code.

\section{Fundamental parameters and weight distributions}
\label{sec:distributionsimplex}

In this section, we calculate the fundamental parameters of the linear simplex codes $\mathcal{S}_{k}^{\alpha}$ and $\mathcal{S}_{k}^{\beta}$  over a finite chain ring $R$, namely their number of codewords and their types, as well as their weight distributions for the Hamming and homogeneous weights. These results are a generalization of the ones given in \cite{Gupta} for $\mathcal{S}_{k}^{\alpha}$ and $\mathcal{S}_{k}^{\beta}$ when $R=\Z_{2^{s}}$. 
The Hamming weight distribution of $\mathcal{S}_{k}^{\beta}$ over a finite chain ring, given in Theorem \ref{theo:HWESimplexBeta}, was previously proved in \cite{HonoldLandjev} using projective Hjelmslev geometries. However, we provide an original proof for this result from a completely different perspective using combinatorial techniques.

\begin{proposition} \label{prop:3.1}
The linear simplex $\alpha$ code $\mathcal{S}_{k}^\alpha$ over $R$ has type $(q^{sk}; k, 0,\dots,0)$.  
\end{proposition}

\begin{proof}
By Definition \ref{def:simplexalpha}, $\cS_k^\alpha$ is generated by $G_k^{\alpha}$, whose columns are all the elements of $R^{k}$. Therefore, its length is $|R^k|=q^{sk}$. Since every element of $R^{k}$ is a column of $G_k^\alpha$, we can permute the columns so that the resulting matrix is in standard form, that is, with $I_k$ in its first columns. Hence, $\cS_k^\alpha$ is permutation equivalent to a code with such a generator matrix. By the definition of type given after Equation (\ref{eq:salagean matrix}), $(t_1,\dots,t_s)=(k,0,\dots,0)$. 
\end{proof}

\begin{proposition} \label{prop:3.3}
The linear simplex $\beta$ code $\mathcal{S}_{k}^{\beta}$ over $R$ has type $(q^{(s-1)(k-1)}\frac{q^{k}-1}{q-1}; k,0,\dots,0).$   
\end{proposition}

\begin{proof}
The length of $\mathcal{S}_{k}^{\beta}$ is given by Proposition \ref{prop:3.2}. By Definition \ref{def:2.5}, $\cS_k^\beta$ is generated by $G_k^{\beta}$. By induction, we see that the columns of $G_k^{\beta}$ can be permuted so that the identity matrix $I_k$ appears in its first columns. First, the matrix $G_1^\beta=(1)$ has the identity matrix $I_1$. Suppose that $G_{k-1}^\beta$ contains the identity matrix $I_{k-1}$. By the recursive construction, we may consider the matrix $\bar{I}_k$ of size $k \times (k-1)$ having all-zeros in the first row and the identity matrix $I_{k-1}$ in the other rows. The first column of $G_{k}^{\beta}$ is always $(1,0,\dots,0)^{T}$. Therefore, we can permute the columns so that the resulting matrix is in standard form, that is, with $I_k$ in its first columns.  
Hence, $\cS_k^\beta$ is permutation equivalent to a code with such a generator matrix. By the definition of type given after Equation (\ref{eq:salagean matrix}), $(t_1,\dots,t_s)=(k,0,\dots,0)$. 
\end{proof}

In order to determine the weight distributions for the linear simplex codes $\mathcal{S}_{k}^{\alpha}$ and $\mathcal{S}_{k}^{\beta}$, we introduce the concept of valuation of elements in $R$ and some remarks related to ideals in $R$. 

For $x\in R \backslash \{0\}$, we define the {\it valuation} of $x$, denoted by $\nu(x)$, as the maximum natural number $m$ such that $x=\gamma^{m}\beta$ for a unit $\beta\in R^{*}$, so $\nu(x)\in \{0,\dots,s-1\}$.  If $x=0$, we formally set $\nu(x)=\infty$. Since any generator of $\langle \gamma \rangle$ is of the form $\gamma \beta$ for some unit $\beta\in R^{*}$, $\nu$ does not depend on the choice of generator of $\langle \gamma \rangle$. In other words, it is  associated with the chain ring itself rather than with a particular generator. Note that we can define a sum  operation ``+" over the valuation set $\Gamma=\{0,1,\dots,s-1\}\cup\{\infty\}$ in a natural way: for $a,b\in \Gamma$, define $a+b=\infty$ if  $a=\infty$, or $b=\infty$, or if $a+b\geq s$ when summed as integers; otherwise, define $a+b$ as the usual integer sum.  The elements of $ \Gamma \backslash \{\infty\}$ are ordered as integers and $a<\infty$ for any $a\in \Gamma \backslash \{\infty\}$. 

From the definition of $\nu$, two basic properties immediately follow. For $x\in R$, $\nu(x)=\infty$ if and only if $x=0$, and $\nu(x)=0$ if and only if $x\in R^{*}$. Moreover, 
it is easy to see that for  $x, y \in R$, $\nu(xy)=\nu(x)+\nu(y)$, 
 $\nu(x+y)\geq \min\{\nu(x),\nu(y)\}$ with equality when $\nu(x)\neq\nu(y)$, and 
 $\nu(x)\leq \nu(y)$ if and only if $x\mid y$.

The valuation $\nu$ can be extended to elements in $R^{n}$ as follows. Given $x=(x_{1},\dots,x_{n})\in R^n$, $\nu(x)=\min\{\nu(x_{1}),\dots,\nu(x_{n})\}$.  This function is also called valuation and denoted by $\nu$. It has similar properties to those of the valuation function for $R$, that is, for $x\in R^n$, $\nu(x)=\infty$ if and only if $x=\zero$, and $\nu(x)=\zero$ if and only if there is some $i\in\{1,\dots,n\}$ such that $x_{i}\in R^{*}$. Moreover,  for $a\in R$ and $x\in R^{n}$, $\nu(ax)=\nu(a)+\nu(x)$, and for $x,y\in R^{n}$, $\nu(x+y)\geq \min\{\nu(x),\nu(y)\}$ with equality when $\nu(x)\neq\nu(y)$.

\begin{remark}\label{rmk:annihilator}
Given $x,y\in R\setminus \{0\}$, $x\in\langle \gamma^{j} \rangle$ for some $j\in\{1,\dots,s-1\}.$ Then, $xy=0$ if and only if $y\in\langle \gamma^{s-j} \rangle.$
\end{remark}

\begin{remark} \label{remark:sumpermv}  
Let $\lambda \in R$ and $j\in\{0,1,\dots,s-1\}$ such that $\nu(\lambda)\geq j$. Then, $\langle \gamma^{j} \rangle +\lambda =\langle \gamma^{j}\rangle.$ In particular, $R+\lambda=R$.
\end{remark}

Now, we determine the weight distributions for the linear simplex codes $\mathcal{S}_{k}^{\alpha}$ and $\mathcal{S}_{k}^{\beta}$, for both the Hamming and homogeneous weights. Due to the fact that the Gray map $\Phi_{R}$ is an isometry using the homogeneous metric in $R$ and the Hamming metric in $\Phi_{R}(R)$, the Hamming weight distributions of $\Phi_{R}( \mathcal{S}_{k}^{\alpha})$ and $\Phi_{R}( \mathcal{S}_{k}^{\beta})$ are also obtained. 
Note that the codes obtained after applying this Gray map  satisfy the property that their Hamming weight distribution coincides with their distance distribution, as $d_H( \Phi_R(u),\Phi_R(v))=w_H(\Phi_R(u-v))$ for any $u,v \in R$ \cite{Greferath}.

We begin by computing the Hamming weight distribution for $\mathcal{S}_{k}^{\alpha}$. 
Let $s_{i}^{k}$ be the $i$th row of $G_{k}^{\alpha}$, $i\in \{1,\ldots, k\}$. Clearly, by construction, all rows of $G_{k}^{\alpha}$ contain the same coordinates but permuted, and $w_{H}(s_{i}^{k})=q^{sk}-q^{s(k-1)}$. Moreover, note that $w_{H}(x)=w_{H}(ux)$, for any $x\in R$ and $u\in R^{*}$. Thus, we know the Hamming weight of every generator $s_i^k$, as well as the Hamming weight of every generator multiplied by an unit of $R$. Before tackling the Hamming weight distribution of $\mathcal{S}_{k}^{\alpha}$, we introduce some results dealing with how multiplying a generator $s_i^k$ by a zero divisor affects its Hamming weight.

\begin{lemma} \label{lemma:3.1}
Let $s_{i}^{k}$ be the $i$th row  of $G_{k}^{\alpha}$, $i\in \{1,\dots,k\}$. For all $j\in\{0,1,\dots,s-1\}$, $\gamma^{j}s_{i}^{k}$ has every element of $\langle \gamma^{j} \rangle$  repeated $q^{j+s(k-1)}$ times.
\end{lemma}

\begin{proof}
By Proposition \ref{prop:2.1}, since all rows $s_{i}^{k}$ are equal up to a permutation, we can prove the result by just considering $s_{k}^{k}=(\rho_{0},\rho_{1},\dots,\rho_{q^{s}-1})^{(q^{s(k-1)})}$. Let $v=(\rho_{0},\rho_{1}\dots,\rho_{q^{s}-1})$. Then, we have that $\gamma^{j}v$ has every element of $\langle \gamma^{j} \rangle$ repeated $q^{j}$ times in its entries. This is a consequence of Remarks \ref{rmk:idealcount} and \ref{rmk:annihilator}, as $\gamma^{j}x=\gamma^{j}y$ if and only if $x-y\in\langle \gamma^{s-j} \rangle.$ In consequence, $\gamma^{j} s_{k}^{k}$ has every element of $\langle \gamma^{j} \rangle$ repeated $q^{j}q^{s(k-1)}$ times in its entries, which concludes the proof.
\end{proof}

\begin{corollary} \label{corollary:2.1}
Let $s_{i}^{k}$ be the $i$th row  of $G_{k}^{\alpha}$, $i\in \{1,\dots,k\}$. For all $j\in\{0,1,\dots,s-1\}$, $w_{H}(\gamma^{j}s_{i}^{k})=q^{sk}-q^{j+s(k-1)}.$
\end{corollary}

Now, we see that studying the Hamming weight of the codewords given in Lemma \ref{lemma:3.1} is enough, as all codewords of $\mathcal{S}_{k}^{\alpha}$ are equal to one of those up to a permutation. We start giving a lemma. 



\begin{lemma} \label{lemma:3.2}
Let $C_R$ be a linear code over $R$ of type $(n;k,0,\dots,0)$ with generator matrix $G$. Let $\{u_{1},\dots,u_{k}\}$ be the row vectors of $G$ and $c\in C_R$. If $c=\sum_{i=1}^{k}{\alpha_{i}u_{i}}$ for some $\alpha_{i}\in R$, then $\nu(c)=\min\{\nu(\alpha_{1}),\dots,\nu(\alpha_{k})\}$.
\end{lemma}

\begin{proof}
Let $\eta=\min\{\nu(\alpha_{1}),\dots,\nu(\alpha_{k})\}$. For $c=\zero$, as $C_R$ is free, $c=\sum_{i=1}^{k}{\alpha_{i}u_{i}}=\zero$ if and only if $\alpha_{i}=0$ for all $i\in \{1,\ldots,k\}$. In consequence, $\nu(\alpha_{i})=\infty$ for all $i\in\{1,\dots,k\}$, so $\eta=\infty=\nu(0)$ and the result is true. 

For $c\neq \zero$, we have that $\nu(c)\in\{0,1,\dots,s-1\}$ and $c=\sum_{i=1}^{k}{\alpha_{i}u_{i}}$ with at least one nonzero $\alpha_{i}$. By definition of $\eta$, we can write every coefficient $\alpha_i$ as $\alpha_{i}=\gamma^{\eta}\beta_{i}$, where $\beta_{i}\in R$. Therefore, $c=\gamma^{\eta}(\sum_{i=1}^{k}{\beta_{i}u_{i}})$, which implies that  $\nu(c)\geq \eta$. 

By the definition of $\nu(c)$, we have that $\gamma^{s-\nu(c)}c=\zero$. Again, as $C_R$ is free and  $c=\sum_{i=1}^{k}{\alpha_{i}u_{i}}$, we obtain that $\gamma^{s-\nu(c)}\alpha_{i}=0$ for all $i\in\{1,\dots,k\}$. In particular, this is true for some index $i_{0}$ such that $\nu(\alpha_{i_{0}})=\eta$. Thus, $\gamma^{s-\nu(c)}\alpha_{i_{0}}=0$, which implies that $\nu(\alpha_{i_{0}})\geq \nu(c)$ by Remark \ref{rmk:annihilator}, so $\eta\geq\nu(c)$. 
\end{proof}

\begin{proposition} \label{prop:2.7}
Let $c\in \mathcal{S}_{k}^{\alpha}\setminus\{\zero\}$ and $s_{i}^{k}$ be the $i$th row  of $G_{k}^{\alpha}$, $i\in \{1,\dots,k\}$. Then, $c$ is equal to $\gamma^{\nu(c)}s_{i}^{k}$, up to a permutation of coordinates, for all $i\in \{1,\dots,k\}$.
\end{proposition}
\begin{proof}
By Lemma \ref{lemma:3.1}, it suffices to prove that $c$ is equal to $\gamma^{\nu(c)} s_i^k$ up to a permutation for some $i\in  \{1,\dots,k\}$, since  all rows of $G_k^\alpha$ contain the same elements, but permuted. We have that $c=\sum_{j=1}^r \alpha_{i_j}s_{i_j}^k$ for some $1\leq r\leq k$ and $1\leq i_1< i_2<\dots < i_r \leq k$, where $0\not =\alpha_{i_j}=\gamma^{n_{i_j}}\beta_{i_j}$  and $n_{i_j}=\nu(\alpha_{i_j})$. Note that $\nu(c)\in \{0,1,\ldots,s-1\}$ since $c\not = \zero$. Moreover, $\nu(c)=\min\{n_{i_1},\dots, n_{i_r}\}$ by Lemma \ref{lemma:3.2}. Then, we can write 
\begin{equation*}\label{eq:expCodeword}
c=\sum_{j=1}^{r}{\gamma^{n_{i_{j}}}\beta_{i_{j}}s_{i_{j}}^{k}}=\gamma^{\nu(c)}(\sum_{j=1}^{r}{\gamma^{n_{i_{j}}-\nu(c)}\beta_{i_{j}}s_{i_{j}}^{k}})=\gamma^{\nu(c)}c',
\end{equation*}
with $\nu(c')=0$. If the statement is true for $c'$, then the result also follows for $c \not =\zero $.  Therefore, from now on, we assume that $\nu(c)=0$.

We prove the result by induction over $k\geq 1$. For $k=1$, there is one summand in (\ref{eq:expCodeword}), so the result is trivially true. Now, assume that the result holds for $k-1$, and consider the general case $k$.

First, assume that $i_1>1$.  By the construction of $G_k^{\alpha}$, $c=(\sum_{j=1}^{r}{\gamma^{n_{i_{j}}}\beta_{i_{j}}s_{i_{j-1}}^{k-1}})^{(q^{s})}=\bar{c}^{(q^{s})}$, for $\bar{c}=\sum_{j=1}^{r}{\gamma^{n_{i_{j}}}\beta_{i_{j}}s_{i_{j-1}}^{k-1}}\in \cS_{k-1}^\alpha$. Therefore, by the induction hypothesis, and since $\nu(\bar{c})=\nu(c)=0$, we conclude that $\bar{c}=s_{1}^{k-1}$, up to a permutation of its coordinates. Hence, by the recursive construction of $G_{k}^{\alpha}$, we have that $c=s_{2}^{k}$ up to a permutation of coordinates.

Now, we assume that  $i_{1}=1$.
In this case,  $$c=
\gamma^{n_{1}}\beta_{1}s_{1}^{k}+\sum_{j=2}^{r}{\gamma^{n_{i_{j}}}\beta_{i_{j}}s_{i_{j}}^{k}}=\gamma^{n_{1}}\beta_{1}s_{1}^{k}+(\sum_{j=2}^{r}{\gamma^{n_{i_{j}}}\beta_{i_{j}}s_{i_{j-1}}^{k-1}})^{(q^{s})}.$$ Since $s_{1}^{k}=(\boldsymbol{\rho_{0}}^{(q^{s(k-1)})},\dots,\boldsymbol{\rho_{q^{s}-1}}^{(q^{s(k-1)})})$ by Proposition \ref{prop:2.1}, $c=(c_0, c_1, \dots, c_{q^s-1})$, where $c_{\ell}=\bar c +\gamma^{n_{1}}\beta_{1}\rho_\ell \mathbf{1}^{(q^{s(k-1)})}$  and 
 $\bar c=\sum_{j=2}^{r}{\gamma^{n_{i_{j}}}\beta_{i_{j}}s_{i_{j-1}}^{k-1}}$ for any $\ell \in \{0,\dots,q^s-1\}$. By induction hypothesis, we know  that $\bar c$ is equal to $\gamma^{\nu(\bar c)}s_{1}^{k-1}$ up to a permutation. By the definition of the valuation, $\nu(c)=\min\{n_{1},\nu(\bar c)\}$, so either $\nu(\bar c)=0$ or $n_{1}=0$. This allows us to divide the remainder of the argument in two different cases.

If $\nu(\bar c)=0$, then $\bar c$ is equal to $s_{1}^{k-1}$ up to a permutation, so $\bar c$ has every element of $R$ repeated $q^{s(k-2)}$ times. By Remark \ref{remark:sumpermv}, the vectors $c_{\ell}=\bar c+\gamma^{n_{1}}\beta_{1}\rho_\ell \mathbf{1}^{(q^{s(k-1)})}
$ have as well all elements of $R$ repeated  $q^{s(k-2)}$ times for any $\ell \in \{0,\dots,q^s-1\}$, so they are equal  to $s_{1}^{k-1}$ up to a permutation, making $c$ equal to $s_{2}^{k}$ up to a permutation. 

Finally, if $\nu(\bar c)>0$, then we have that $n_{1}=0$. 
Since $\beta_1$ is a unit, up to a permutation, $c$ is equal to $c'=(c'_0,c'_1,\dots, c'_{q^s-1})$, where $c'_\ell =\bar c + \rho_\ell \mathbf{1}^{(q^{s(k-1)})}$. We can assume that $\bar c \not =\zero$, otherwise the result follows. 
By induction hypothesis, we know that $\bar c$ is equal to $\gamma^{\nu(\bar c)}s_{1}^{k-1}$ up to a permutation, so $\bar c$ has every element of $\langle \gamma^{\nu(\bar c)}\rangle $ repeated $q^{\nu(\bar c)+s(k-2)}$ times by Lemma \ref{lemma:3.1}. 
Let $I=\langle \gamma^{\nu(\bar c)}\rangle $ and let $v_I$ be a vector having exactly all  the elements of $I$ in its coordinates. Note that the vector $v=(v_I, v_I+\boldsymbol{\rho_1}, \ldots, v_I+\boldsymbol{\rho_{q^s-1}})$ contains every element of $R$ exactly $|I|=q^{s-\nu(\bar{c})}$ times. Indeed, this follows from the fact that $R+\lambda=R$ for every $\lambda\in I$, as established in Remark \ref{remark:sumpermv}. Moreover, since $\bar{c}$ is equal, up to a permutation, to $\gamma^{\nu(\bar{c})}s_1^{k-1}$, we can suitably reorder the coordinates of $c'$ to obtain a vector consisting of $q^{\nu(\bar{c})+s(k-2)}$ repetitions of $v$. Consequently, every element of $R$ occurs in $c'$ exactly $ q^{s-\nu(\bar{c})}q^{\nu(\bar{c})+s(k-2)}=q^{s(k-1)}$ times. Therefore, $c'$ is equal to $s_1^k$, up to a permutation of its coordinates.
\end{proof}

\begin{corollary} \label{coro:HammingWeightCodeword}
Let $c\in \mathcal{S}_{k}^{\alpha}\setminus \{\zero\}$. Then, $w_{H}(c)=q^{sk}-q^{\nu(c)+s(k-1)}.$    
\end{corollary}

\begin{proof}
It follows from Proposition \ref{prop:2.7} and Lemma \ref{lemma:3.1}.
\end{proof}

\begin{corollary}\label{coro:MinHamWeightSimplexAlpha}
The minimum Hamming distance of $\mathcal{S}_{k}^{\alpha}$  is $d_{H}(\mathcal{S}_{k}^{\alpha})=(q-1)q^{sk-1}.$
\end{corollary}

By Corollary \ref{coro:HammingWeightCodeword}, the Hamming weight of any nonzero codeword $c\in \mathcal{S}_{k}^{\alpha}$ is determined by its valuation $\nu(c) \in \{0,1,\dots,s-1\}$. We now compute the number of codewords corresponding to each valuation $j\in \{0,1,\dots,s-1\}$; that is, for each possible nonzero Hamming weight $w_j=q^{sk}-q^{j+s(k-1)}$, thereby completing the study of the Hamming weight distribution for the linear simplex $\alpha$ code $\mathcal{S}_{k}^\alpha$. First, we count the number of codewords of each valuation $j\in\{0,1,\dots,s-1\}$ in any free code over a finite chain ring $R$.


\begin{lemma}\label{Lemma:valuationFree}
Let $R$ be a finite chain ring and $C_R$ a linear code over $R$ of type $(n;k,0,\dots,0)$ with generator matrix $G$ as in (\ref{eq:salagean matrix}). Let $j\in\{0,1,\dots,s-1\}$. The number of codewords of $C_R$ with valuation equal to $j$ is $q^{k(s-j)}-q^{k((s-j)-1)}$.
\end{lemma}

\begin{proof}
Let $s_1,\dots, s_k$ be the row vectors of $G$. We have that any codeword $c\in C_R$ can be expressed as $c=\sum_{i=1}^{k}{\alpha_i s_i}$ for some $\alpha_i\in R$. Moreover, $\nu(c)=\min\{{\nu(\alpha_1),\dots,\nu(\alpha_k)}\}$ by Lemma \ref{lemma:3.2}. Applying Remark \ref{rmk:idealcount}, we have that the number of codewords with valuation at least $j$ is equal to $|\langle \gamma^j  \rangle|^k=q^{k(s-j)}$. Note that codewords with valuation $j$ are exactly those having valuation at least $j$ but not $j+1$. If  $j \in \{0,\dots,s-2\}$, there are $q^{k(s-j)}-q^{k(s-j-1)}$ of them. If $j=s-1$, since there is only one codeword with  valuation greater than $s-1$, the zero codeword, the statement is satisfied since $q^{k(s-(s-1)-1)}=1$.
\end{proof}


\begin{theorem}\label{theo:HWESimplexAlpha}
The Hamming weight enumerator for $\cS_{k}^{\alpha}$ is 

$$W_{\cS_{k}^{\alpha}}(X,Y)=\sum_{j=0}^{s-1}{A_{w_j}X^{q^{sk}-w_j}Y^{w_j}}+X^{q^{sk}},$$ 
where $w_j=q^{sk}-q^{s(k-1)+j}$ and $A_{w_j}=q^{k(s-j)}-q^{k((s-j)-1)}$.  
\end{theorem}

\begin{proof}
Let $c\in \cS_{k}^{\alpha}$. By Corollary \ref{coro:HammingWeightCodeword}, either $c=\zero$ and $w_H(c)=0$, or $c\not=\zero$ and $w_{H}(c)=q^{sk}-q^{\nu(c)+s(k-1)}.$ We have that $A_0=1$. Let $A_{w_j}$ be the number of codewords of weight $w_j=q^{sk}-q^{j+s(k-1)}$ in $\mathcal{S}_{k}^\alpha$, for $j\in \{0,1,\dots, s-1\}$. Since $A_{w_j}=|\{c \in\mathcal{S}_{k}^{\alpha}\mid w_{H}(c)=w_j\}|$, where $w_j=q^{sk}-q^{j+s(k-1)}$, 
we can write that $A_{w_j}=|\{c \in\mathcal{S}_{k}^{\alpha}\mid \nu(c)=j\}|$. By Lemma \ref{Lemma:valuationFree}, $A_{w_j}= q^{k(s-j)}-q^{k((s-j)-1)}$ and the statement follows.
\end{proof}

Note that summing all the elements of the Hamming weight distribution of $\mathcal{S}_{k}^\alpha$ given by Theorem \ref{theo:HWESimplexAlpha}, we obtain the total number of codewords, $|\mathcal{S}_{k}^\alpha |=q^{sk}$. That is, $\sum_{i=0}^n A_i = A_0 + \sum_{j=0}^{s-1} A_{w_j}=q^{sk}$.  Indeed, 
$A_0=1$ and  $\sum_{j=0}^{s-1}{A_{w_j}}=\sum_{j'=1}^{s}{q^{kj'}-q^{k(j'-1)}}$, where $j'=s-j$. This is a telescoping sum, so we have that $\sum_{j=0}^{s-1}{A_{w_j}}=q^{sk}-1$. 

Since the study of the Hamming weight distribution has provided extensive information about the codewords of $\mathcal{S}_{k}^{\alpha}$, such as which elements they have and how many times each of them appears, we can also calculate the homogeneous weight of any codeword as well as the homogeneous weight distribution of $\mathcal{S}_{k}^{\alpha}$. 

\begin{proposition} \label{prop:HomWeightCodeword}
Let $c\in \mathcal{S}_{k}^{\alpha}\setminus \{\zero \}$. Then, $w_{Hom}(c)=q^{s(k+1)-2}(q-1)$.  
\end{proposition}
\begin{proof}
To determine the homogeneous weight of a nonzero codeword $c\in \mathcal{S}_{k}^{\alpha}$, we have to count how many coordinates of $c$ are zero, and how many of those that are not zero are in $\langle \gamma^{s-1} \rangle$. 
By Proposition \ref{prop:2.7}, it suffices to consider the codewords $c_j=\gamma^{j}s_{k}^{k}$, with $j\in \{0,1,\ldots, s-1\}$. By Lemma \ref{lemma:3.1}, we know that $c_j$ has $q^{j+s(k-1)}$ zero coordinates, which have weight zero and do not count. We can easily see that $c_j$ has $(q-1)q^{j+s(k-1)}$ nonzero coordinates that belong to $\langle \gamma^{s-1} \rangle$, as $|\langle \gamma^{s-1}\rangle|=q$. These coordinates have homogeneous weight equal to $q^{s-1}$. Finally, the remaining $(q^{s-j}-q)q^{j+s(k-1)}$ coordinates have homogeneous weight equal to $(q-1)q^{s-2}$, because $|\langle \gamma^{j} \rangle \setminus \langle \gamma^{s-1} \rangle|=q^{s-j}-q$. Therefore, we have that 
\begin{equation*} \begin{split}    
w_{Hom}(c_j)&=(q-1)q^{j+s(k-1)}q^{s-1}+(q^{s-j}-q)q^{j+s(k-1)}(q-1)q^{s-2}\\
&=q^{j+s(k-1)}(q-1)(q^{s-1}+(q^{s-j}-q)q^{s-2})\\
&=q^{j+s(k-1)}(q-1)q^{2s-j-2}=q^{s(k-1)+2s-2}(q-1)=q^{s(k+1)-2}(q-1).
\end{split}
\end{equation*}
Note that $w_{Hom}(c_j)$ does not depend on $j$, so we conclude that all nonzero codewords have the same homogeneous weight. 
\end{proof}

\begin{corollary} \label{cor:minhomalpha}
The minimum homogeneous weight of $\mathcal{S}_{k}^{\alpha}$  is $d_{Hom}(\mathcal{S}_{k}^{\alpha})=q^{s(k+1)-2}(q-1).$
\end{corollary}

Recall that $\Phi_{R}:R\longrightarrow \F_q^{q^{s-1}}$  is an isometry that maps the homogeneous weight of an element to the Hamming weight of its image \cite{Greferath}. Therefore, we can deduce the fundamental parameters and the Hamming weight distribution of the $R$-linear code $\Phi_{R}(\cS_k^{\alpha})$ over $\F_q$. Moreover, since $R$-linear codes are distance-invariant \cite{Greferath}, this also determines the Hamming distance distribution of $\Phi_{R}(\cS_k^{\alpha})$.


\begin{theorem} \label{theo:simplexAlphaParameters}
The $R$-linear code $C=\Phi_{R}(\mathcal{S}_{k}^\alpha)$ is a $(q^{s(k+1)-1}, q^{sk}, q^{s(k+1)-2}(q-1))$ code over $\mathbb{F}_{q}$. Moreover, the Hamming weight enumerator for $C$  is $W_{C}(X,Y)=A_{\ell}X^{N-\ell}Y^{\ell} + X^{N},$ where $N=q^{s(k+1)-1}$, $A_\ell=q^{sk}-1$  and $\ell=q^{s(k+1)-2}(q-1).$
\end{theorem}

\begin{proof}
The length of $\Phi_{R}(\mathcal{S}_{k}^{\alpha})$ is the length of $\mathcal{S}_{k}^{\alpha}$ multiplied by $q^{s-1}$, so $N=q^{s-1}n=q^{s-1}q^{sk}=q^{s(k+1)-1}$ by Proposition \ref{prop:3.1}. The number of codewords is $|\Phi_{R}(\mathcal{S}_{k}^{\alpha})|=|\mathcal{S}_{k}^{\alpha}|=q^{sk}$, since $\Phi_{R}$ is injective and $\mathcal{S}_{k}^{\alpha}$ is a free $R$-module.

We know that the Hamming weight of the codewords of $\Phi_{R}(\mathcal{S}_{k}^{\alpha})$ is equal to the homogeneous weight of the codewords of $\mathcal{S}_{k}^{\alpha}$. By Proposition \ref{prop:HomWeightCodeword},  for $c\in\Phi_{R}(\mathcal{S}_{k}^{\alpha})$, $w_{H}(c)=0$ if $c=0$ and $w_{H}(c)=q^{s(k+1)-2}(q-1)$ otherwise. The minimum Hamming distance of $\Phi_{R}(\mathcal{S}_{k}^{\alpha})$ is the minimum homogeneous weight of $\mathcal{S}_{k}^{\alpha}$, which is equal to $q^{s(k+1)-2}(q-1)$ by Corollary \ref{cor:minhomalpha}.
\end{proof}

Table \ref{table:SimplexAlpha} summarizes the Hamming and homogeneous weight distributions of $\cS_k^{\alpha}$. 

\begin{table}[ht]
    \centering
    \begin{tabular}{|c|c|c|c|}
        \hline
         Valuation $\nu(c)$ &$w_{H}(c)$ &$w_{Hom}(c)$ &Number of codewords  \\
         \hline 
         $0\leq j\leq s-1$ &$q^{sk}-q^{j+s(k-1)}$  &$q^{s(k+1)-2}(q-1)$ &$q^{k(s-j)}-q^{k(s-j-1)}$\\
         \hline
         $\infty$ &0 &0 &1\\
         \hline
    \end{tabular}
    \caption{ Hamming and homogeneous weight distributions of $\cS_k^{\alpha}$ in terms of the valuation of its codewords.}
    \label{table:SimplexAlpha}\end{table}   
    
\vspace{-0.75truecm}



\begin{remark}
Consider the case $s=1$, so that $R=\F_q$. In this setting, the homogeneous weight and the Hamming weight coincide, and $\Phi_R$ is the identity map. By Theorem~\ref{theo:simplexAlphaParameters}, $\cS_k^{\alpha}$ is a linear code over $\F_q$ of length $q^k$ and dimension $k$, in which every nonzero codeword has Hamming weight $q^{k-1}(q-1)$. Hence, $\cS_k^{\alpha}$ is a $[q^k,k,q^{k-1}(q-1)]$ $q$-ary linear code. Moreover, from the construction of its generator matrix given in Definition~\ref{def:simplexalpha}, we have $\cS_k^{\alpha}$ is the subcode of the first-order Reed-Muller code over $\F_q$  obtained by removing the  row of its generator matrix whose entries are all equal to $1$.
\end{remark}

Finally, we present several detailed examples for small values of $q$, $s$, and $k$, together with the  Hamming and homogeneous weight distributions.

\begin{example}
Let $R=\Z_{4}$ and $k=2$. In this case, $q=s=2$, and the homogeneous weight coincides with the Lee weight. By Definition \ref{def:simplexalpha}, the generator matrix of the linear simplex $\alpha$ code  $\mathcal{S}_2^\alpha$ over $\Z_4$ is
$$G_{2}^{\alpha}=\left(\begin{array}{cccc}
\mathbf{0} &\mathbf{1} &\mathbf{2} &\mathbf{3}\\
0\,1\,2\,3 &0\,1\,2\,3 &0\,1\,2\,3 &0\,1\,2\,3
\end{array}
\right)=\left(\begin{array}{c}
s_1^{2}\\ 
s_2^{2}\end{array}\right).$$
Therefore, $\cS_2^{\alpha}$ has type $(16;2,0)$, as shown in Proposition \ref{prop:3.1}.


The Hamming weight of a nonzero codeword $c$ is $w_0=2^{4}-2^{2}=12$ or $w_1=2^{4}-2^{2+1}=8$, depending on whether $c$ has valuation $0$ or $1$, respectively. For example, $w_H(s_1^2)=w_H((0000111122223333))=16-4=12$ and $w_H(2s_2^2)=w_H((0202020202020202))=8$. Since there are $A_{w_0}=2^4-2^2=12$ codewords with valuation $0$ and $A_{w_1}=2^2-1=3$ codewords with valuation $1$,
the Hamming weight enumerator of $\cS_2^{\alpha}$ is  $W_{\cS_2^\alpha}(X,Y)=12X^{4}Y^{12} + 3X^{8}Y^{8}+X^{16}$ as shown in Theorem \ref{theo:HWESimplexAlpha}. 

The homogeneous weight of every nonzero codeword is $(2-1)2^{2(2+1)-2}=2^{4}=16$. Indeed, the elements $1$ and $3$ of $\Z_4$, which have valuation $0$, have homogeneous weight $1$, whereas the element $2$, with valuation $1$, has homogeneous weight $2$.  For example,
$w_{Hom}(s_1^{2})=8\cdot 1+4\cdot 2=16$ and $w_{Hom}(2s_2^{2})=8\cdot 2=16$. The homogeneous weight enumerator of $\cS_2^\alpha$ coincides with the Hamming weight enumerator of its Gray map image $C=\Phi_R(\cS_2^{\alpha})$, which is $W_{C}(X,Y)=15X^{16}Y^{16} + X^{32}$. Hence, $C$ is a binary  $(32,16,16)$ code  as shown in Theorem \ref{theo:simplexAlphaParameters}.


\end{example}

\begin{example}
Let $R=\Z_{8}$ and $k=2$. In this case, $q=2$ and $s=3$. By Definition \ref{def:simplexalpha}, the generator matrix of the linear simplex $\alpha$ code  $\mathcal{S}_2^\alpha$ over $\Z_8$ is
$$G_{2}^{\alpha}=\left(\begin{array}{cccc}
\mathbf{0} &\mathbf{1} &\dots &\mathbf{7}\\
0\,1\,2\,3\,4\,5\,6\,7 &0\,1\,2\,3\,4\,5\,6\,7 &\dots &0\,1\,2\,3\,4\,5\,6\,7
\end{array}
\right)=\left(\begin{array}{c}
s_1^{2}\\ 
s_2^{2}\end{array}\right).$$
Hence, $\mathcal{S}_2^\alpha=\{ as_1^2 + bs_2^2 \mid a,b \in \Z_8\}$ has type $(64;2,0)$, as stated in Proposition \ref{prop:3.1}. 

The Hamming weight of a nonzero codeword $c$ is $w_0=2^6-2^3=56$, $w_1=2^6-2^4=48$, or $w_2=2^6-2^5=32$, depending on whether $\nu(c)$ is $0$, $1$, or $2$, respectively. Since $A_{w_0}=2^6-2^4=48$, $A_{w_1}=2^4-2^2=12$, and $A_{w_2}=2^2-1=3$, the Hamming weight enumerator of $\cS_2^{\alpha}$ is  $W_{\cS_2^\alpha}(X,Y)=48X^{8}Y^{56} + 12X^{16} Y^{48} +3X^{32}Y^{32}+X^{64}$ as shown in Theorem \ref{theo:HWESimplexAlpha}. 

The homogeneous weight of every nonzero codeword is $(2-1)2^{3(2+1)-2}=128$. Indeed, the element $4\in \Z_8$ has homogeneous weight $4$, and the remaining nonzero elements have homogeneous weight $2$.  Then, for example,
$w_{Hom}(s_1^{2})=8\cdot 6 \cdot 2+8\cdot 4=128$, $w_{Hom}(2s_2^{2})=w_{Hom}(0,2,4,6,\dots,0,2,4,6)=8\cdot 4 \cdot 2 + 16 \cdot 4=128$, and $w_{Hom}(4s_2^{2})=w_{Hom}(0,4,0,4,\dots,0,4)=32\cdot 4=128$. The homogeneous weight enumerator of $\cS_2^\alpha$ coincides with the Hamming weight enumerator of its Gray map image $C=\Phi_R(\cS_2^{\alpha})$, which is $W_{C}(X,Y)=255X^{128}Y^{128}+X^{256}$. Hence, $C$ is a binary  $(256,64,128)$ code as shown in Theorem \ref{theo:simplexAlphaParameters}.
\end{example}

\begin{example} Let $R=GR(4,2)$, as in Example \ref{ex:SimplexAlphaGR42}, and $k=1$. In this case, $q=4$ and $s=2$. The generator matrix of the linear simplex $\alpha$ code $\cS_1^{\alpha}$ over $GR(4,2)$ is $G_1^\alpha = (s_1^1)$, where $s_1^1=(0,1,\omega,\omega+1,2,3,2+\omega,3+\omega,2\omega,2\omega+1,3\omega,3\omega+1,2\omega+2,2\omega+3,3\omega+2,3\omega+3)$. Hence, $\cS_1^{\alpha}=\{as_1^1\mid a\in GR(4,2)\}$ has type $(16;1,0)$, as stated in Proposition \ref{prop:3.1}.

The Hamming weight of a nonzero codeword $c$ is $w_0=4^2-1=15$ or $w_1=4^{2}-4=12$, depending on whether $\nu(c)$ is equal to $0$ or $1$, respectively. Since $A_{w_0}=4^{2}-4=12$ and $A_{w_1}=4-1=3$, the Hamming weight enumerator of $\cS_1^{\alpha}$ is $W_{\cS_1^{\alpha}}(X,Y)=12XY^{15}+3X^{4}Y^{12}+X^{16}$, as shown in Theorem \ref{theo:HWESimplexAlpha}.

The homogeneous weight of every nonzero codeword is $(4-1)4^{2(1+1)-2}=48$. Indeed, for $R=GR(4,2)$, elements $2, 2\omega$, and $2\omega+2$ have homogeneous weight $4$, while the rest of nonzero elements have homogeneous weight $3$. Then, for example, $w_{Hom}(s_1^1)=12\cdot3+3\cdot 4=48$ and $w_{Hom}(2s_1^1)=12\cdot 4=48$. The homogeneous weight enumerator of $\cS_1^{\alpha}$ coincides with the Hamming weight enumerator of its Gray map image, $C=\Phi_R(\cS_1^{\alpha})$. Therefore, $W_{C}(X,Y)=15X^{16}Y^{48}+X^{64}$, and $C$ is a $(64,16,48)$ code over $\F_4$, as shown in Theorem \ref{theo:simplexAlphaParameters}.
\end{example}

We now aim to apply the same methods to find the Hamming and homogeneous weight distributions for $\mathcal{S}_{k}^{\beta}$. Since these distributions are trivial for $\mathcal{S}_{1}^{\beta}=R$, we will consider only $k\geq 2$. We begin with some basic observations about the rows of $G_{k}^{\beta}$. Let $h_{i}^{k}$ be the $i${th} row of this matrix, $i\in \{1,\dots,k\}$.
Given $n\geq 1$, we denote as $\mathbf{z}_{(n)}=(a_{0}\boldsymbol{\gamma }^{(n)}, a_{1}\boldsymbol{\gamma }^{(n)},\dots, a_{q^{s-1}-1}\boldsymbol{\gamma }^{(n)})$, a vector listing all elements of $ \langle\gamma \rangle $ repeated $n$ times. Note that $a_{0}\boldsymbol{\gamma }^{(n)}=\zero^{(n)}$, since $a_0=0$. The following remark comes from the recursive construction of $G_{k}^{\beta}$. 

\begin{remark} \label{remark:2.2}
Let $s_{i}^{k}$ and $h_{i}^{k}$ be the $i$th row  of $G_{k}^{\alpha}$ and 
 $G_k^\beta$, respectively, $i\in \{1,\dots,k\}$ and $k\geq 2$. Then, $h_{1}^{k}=(\mathbf{1}^{(q^{s(k-1)})}, \mathbf{z}_{(L_{\beta}(k-1))})$, and $h_{i}^{k}=(s_{i-1}^{k-1},(h_{i-1}^{k-1})^{(q^{s-1})})$ for $i \in \{2,\dots,k\}$.
\end{remark}

\begin{lemma} \label{lemma:2.3}
Let $h_{i}^{k}$ be the $i${th}  row of $G_k^\beta$, $i\in \{1,\dots,k\}$ and $k\geq 2$. Then, $h_{i}^{k}$ has every element of $\langle \gamma \rangle$  repeated $L_{\beta}(k-1)$ times and the remaining $q^{s(k-1)}$ coordinates are in $R^{*}$.
\end{lemma}
\begin{proof}
We show the result by induction over $k\geq 2$. For $k=2$, we have that $h_{1}^{2}=(\mathbf{1}^{(q^{s})}, \mathbf{z}_{(1)})$ and $h_{2}^{2}=(s_{1}^{1},\mathbf{1}^{(q^{s-1})}).$ Since $L_{\beta}(1)=1$, $h_{1}^{2}$ satisfies the proposition trivially. Since every element in $R$ appears exactly once in the coordinates of $s_{1}^{1}$, $h_{2}^{2}$ satisfies the property as well.

Now, suppose it is true for $k-1$.  For $h_{1}^{k}$, the property follows directly from Remark \ref{remark:2.2}. Given $i\in \{2,\dots,k\}$, we know that $h_{i}^{k}=(s_{i-1}^{k-1},(h_{i-1}^{k-1})^{(q^{s-1})})$ by the same remark. 
Applying the induction hypothesis and Lemma \ref{lemma:3.1}, we infer that every element of $\langle \gamma \rangle$ appears $q^{s(k-2)}+q^{s-1}L_{\beta}(k-2)$ times in $h_{i}^{k}$, which equals $L_{\beta}(k-1)$ by (\ref{eq:recsimp}). Finally, we have that there are $q^{s(k-1)}$ coordinates remaining over $R^{*}$, again by equation (\ref{eq:recsimp}), since $|\langle \gamma \rangle|=q^{s-1}$.
\end{proof}

\begin{lemma} \label{lemma:2.4}
Let $c\in R^{n}$, and $\lambda\in R\setminus\{0\}$. If $c$ has every element of $\langle \gamma \rangle$  repeated $m$ times, then $\lambda c$ has every element of $\langle \gamma^{\nu(\lambda)+1}\rangle$  repeated  $q^{\nu(\lambda)}m$ times and the remaining coordinates are in $\langle \gamma^{\nu(\lambda)} \rangle \setminus \langle \gamma^{\nu(\lambda)+1} \rangle$. 
\end{lemma}
\begin{proof}
The coordinates of $\lambda c$ corresponding to the $n-q^{s-1}m$ units of $c$ have valuation equal to $\nu(\lambda)$, so they are in $\langle \gamma^{\nu(\lambda)} \rangle \setminus \langle  \gamma^{\nu(\lambda)+1}\rangle$. The coordinates of $\lambda c$ that correspond to the elements of $\langle \gamma \rangle$ in $c$ have valuation strictly greater than $\nu(\lambda)$, so they are elements of $\langle  \gamma^{\nu(\lambda)+1}\rangle$.
As $\lambda=\gamma^{\nu(\lambda)}u$ for some unit $u$ and $\gamma^{\nu(\lambda)}uz_{1}=\gamma^{\nu(\lambda)}uz_{2}$ if and only if $z_{1}-z_{2}\in \langle \gamma^{s-\nu(\lambda)} \rangle$, each element of $\langle \gamma^{\nu(\lambda)+1} \rangle$ in $\lambda c$ corresponds to  $|\langle \gamma^{s-\nu(\lambda)} \rangle |=q^{\nu(\lambda)}$ elements of $\langle \gamma \rangle$ in $c$. Therefore, if there are $m$ of each element of $\langle \gamma \rangle$ in $c$, there are $q^{\nu(\lambda)}m$ of each element of $\langle \gamma^{\nu(\lambda)+1)} \rangle$ in $\lambda c$.
\end{proof}

\begin{proposition} \label{prop:2.10}
Let  $c\in \mathcal{S}_{k}^{\beta}\setminus \{\mathbf{0}\}$ with $k\geq 2$. Then, $c$ has every element of $\langle \gamma^{\nu(c)+1} \rangle$ repeated $q^{\nu(c)}L_{\beta}(k-1)$ times and the remaining $q^{s(k-1)}$ coordinates are in $\langle \gamma^{\nu(c)} \rangle\setminus \langle \gamma^{\nu(c)+1} \rangle$.   
\end{proposition}
\begin{proof}
Given $c\in\mathcal{S}_{k}^{\beta}$, $c\neq \zero$, we have that $c=\sum_{j=1}^{r}{\gamma^{n_{i_{j}}}\beta_{i_{j}}h_{i_{j}}^{k}}$, where $1\leq i_{1}<i_{2}<\dots<i_{r}\leq k$, $1\leq r\leq k$, and $\nu(\beta_{i_j})=0$ for all $j\in \{1,\dots,r\}$. By Lemma \ref{lemma:3.2}, $\nu(c)=\min{\{n_{i_{1}},\dots,n_{i_{r}}\}}$. Note that
$$c=\gamma^{\nu(c)}\sum_{j=1}^{r}{\gamma^{n_{i_{j}}-\nu(c)}\beta_{i_{j}}h_{i_{j}}^{k}}=\gamma^{\nu(c)}c',$$
where $c'$ is also a codeword and $\nu(c')=0$. Applying Lemma \ref{lemma:2.4} to $c'$, we have that if the result is true for $c'$, then it is true for $c$. Therefore, we can assume that $\nu(c)=0$ for the remainder of the proof.

We prove the result by induction over $k\geq 2$. First, consider $k=2$. By Remark \ref{remark:2.2}, $h_{1}^{2}=(\mathbf{1}^{(q^{s})}, \mathbf{z}_{(1)})$ and $h_{2}^{2}=(s_{1}^{1},\mathbf{1}^{(q^{s-1})})$. Given a codeword $c\in \mathcal{S}_{k}^{\beta}\setminus \{\mathbf{0}\}$ such that $\nu(c)=0$, we show that $c$ has every element of $\langle \gamma\rangle$ exactly once. We have that $c=\gamma^{n_{1}}\beta_{1}h_{1}^{2}+\gamma^{n_{2}}\beta_{2}h_{2}^{2}=(v_1,v_2)$, where $v_{1}=\gamma^{n_{1}}\beta_{1}\mathbf{1}^{(q^{s})}+\gamma^{n_{2}}\beta_{2}s_{1}^{1}$ and $v_{2}=\gamma^{n_{1}}\beta_{1}\mathbf{z}_{(1)}+\gamma^{n_{2}}\beta_{2}\mathbf{1}^{(q^{s-1})}$. We consider two different cases, depending on whether $n_{1}=0$ or $n_{2}=0$.

 If $n_{2}=0$, since $\beta_{2}s_{1}^{1}$ has all elements of $R$ exactly once,  we obtain that $v_{1}=\gamma^{n_{1}}\beta_{1}\mathbf{1}^{(q^{s})}+\beta_{2}s_{1}^{1}$ has every element of $\langle \gamma\rangle$  exactly once by Remark \ref{remark:sumpermv}. Now, for $v_{2}=\gamma^{n_{1}}\beta_{1}\mathbf{z}_{(1)}+\beta_{2}\mathbf{1}^{(q^{s-1})}$, as all coordinates of $\gamma^{n_{1}}\beta_{1}\mathbf{z}_{(1)}$ are in $\langle \gamma \rangle$ and $\nu(\beta_{2})=0$, all coordinates of $v_{2}$ are units. In consequence, $c$ satisfies the desired property.

 If $n_{2}>0$, then $n_{1}=0$. For $v_{1}=\beta_{1}\mathbf{1}^{(q^{s})}+\gamma^{n_{2}}\beta_{2}s_{1}^{1}$, we have that all its coordinates are units. For $v_{2}=\beta_{1}\mathbf{z}_{(1)}+\gamma^{n_{2}}\beta_{2}\mathbf{1}^{(q^{s-1})}$, we have that $\beta_{1}\mathbf{z}_{(1)}$ has every element of $\langle \gamma\rangle$ exactly once, and $\gamma^{n_{2}}\beta_{2}\in \langle \gamma\rangle$ since $n_{2}>0$. Therefore, applying Remark \ref{remark:sumpermv}, $v_{2}$ contains every element of $\langle \gamma \rangle$  exactly once. 
 Then, again, $c$ satisfies the desired property.

Now, consider $k\geq 3$ and that the property is true for $k-1$. Consider $c\in \mathcal{S}_{k}^{\beta}\setminus\{\zero\}$ such that $\nu(c)=0$. We have to prove that $c$ has every element of $\langle \gamma \rangle$  repeated $L_{\beta}(k-1)$ times. We have that $c=\sum_{j=1}^{r}{\gamma^{n_{i_{j}}}\beta_{i_{j}}h_{i_{j}}^{k}}$. We consider two cases taking into account whether $i_1=1$ or $i_1>1$. 
First, suppose that $i_{1}>1$. By Remark \ref{remark:2.2}, $$c=(\sum_{j=1}^{r}{\gamma^{n_{i_{j}}}\beta_{i_{j}}s_{i_{j}-1}^{k-1}},(\sum_{j=1}^{r}{\gamma^{n_{i_{j}}}\beta_{i_{j}}h_{i_{j}-1}^{k-1}})^{(q^{s-1})})=(v_{1},v_{2}).$$ Here, $v_{1}\in S_{k-1}^{\alpha}$ and $\nu(v_1)=0$, so $v_1$ has every element of $R$  repeated $q^{s(k-2)}$ times by Lemma  \ref{lemma:3.1}. In particular, this is true for every element of $\langle \gamma \rangle$. We have that $v_{2}$ is equal to a codeword of $S_{k-1}^{\beta}$ with valuation $0$, repeated $q^{s-1}$ times. Thus, by induction hypothesis, $v_2$ has every element of $\langle \gamma \rangle$ repeated $q^{s-1}L_{\beta}(k-2)$ times. Therefore, $c$ has every element of $\langle \gamma \rangle$ repeated $q^{s(k-2)}+q^{s-1}L_{\beta}(k-2)=L_{\beta}(k-1)$ times by (\ref{eq:recsimp}) and the statement follows.

Now, suppose $i_{1}=1$. In this case, $c=\gamma^{n_{1}}\beta_{1}h_{1}^{k}+c'=(v_{1},v_{2})$, where  $c'=\sum_{j=2}^{r}{\gamma^{n_{i_{j}}}\beta_{i_{j}} h_{i_{j}}^{k} }$, $v_{1}=\gamma^{n_{1}}\beta_{1}\mathbf{1}^{(q^{s(k-1)})}+\sum_{j=2}^{r}{\gamma^{n_{i_{j}}}\beta_{j}s_{i_{j}-1}^{k-1}}$ and $v_{2}=\gamma^{n_{1}}\beta_{1}\mathbf{z}_{L_{\beta}(k-1)}+(\sum_{j=2}^{r}{\gamma^{n_{i_{j}}}\beta_{j}h_{i_{j}-1}^{k-1}})^{(q^{s-1})}$ by Remark \ref{remark:2.2}. 
Since $0=\nu(c)=\min\{n_{1},\nu(c')\}$, at least one of these values is equal to $0$.




If $\nu(c')=0$, it is clear that $v_{1}$ has every element of $R$ repeated $q^{s(k-2)}$ times by Remark \ref{remark:sumpermv} and Lemma \ref{lemma:3.1}. In particular, this is true for every element of $\langle \gamma \rangle$. For $v_{2}$, we can argue the following. We know that $w=\sum_{j=2}^{r}{\gamma^{n_{i_{j}}}\beta_{i_{j}}h_{i_{j}-1}^{k-1}}$ has every every element of $\langle \gamma \rangle$  repeated $L_{\beta}(k-2)$ times, by induction hypothesis and the fact that $\nu(w)=0$. Now, every coordinate of $\gamma^{n_{1}}\beta_{1}\mathbf{z}_{(L_{\beta}(k-1))}$ is in $\langle \gamma \rangle$, and $v_{2}$ is equal to the concatenation of the vectors $v_{z}=w+\mathbf{1}^{(L_{\beta}(k-1))}\gamma^{n_{1}}\beta_{1}z$, for every $z\in \langle \gamma \rangle$. Applying Remark \ref{remark:sumpermv}, we have that $v_{z}$ has every element of $\langle \gamma \rangle$  repeated $L_{\beta}(k-2)$ times. In total, $v_{2}$ has every element of $\langle \gamma \rangle$ repeated $q^{s-1}L_{\beta}(k-2)$ times. The statement follows by (\ref{eq:recsimp}).

Finally, if $\nu(c')>0$, then we have that $n_{1}=0$. Since multiplying by a unit does not modify the number of elements in $\langle \gamma \rangle$, we can assume that $\beta_1=1$. Thus, we have that $c=(v_1,v_2)$  where $v_{1}=\mathbf{1}^{(q^{s(k-1)})}+\sum_{j=2}^{r}{\gamma^{n_{i_{j}}}}\beta_{i_{j}}s_{i_{j}-1}^{k-1}$ and $v_{2}=\mathbf{z}_{L_{\beta}(k-1)}+(\sum_{j=2}^{r}{\gamma^{n_{i_{j}}}\beta_{i_{j}}h_{i_{j}-1}^{k-1}})^{(q^{s-1})}.$ First, note that every coordinate of $\sum_{j=2}^{r}{\gamma^{n_{i_{j}}}}\beta_{i_{j}}s_{i_{j}-1}^{k-1}$ is in $\langle \gamma^{\nu(c')} \rangle$. Note also that $\langle \gamma^{\nu(c')} \rangle \subseteq \langle \gamma \rangle$ since $\nu(c')>0$. The coordinates of $v_{1}$ are units because they are the result of adding a unit to an element of $\langle \gamma \rangle$. Therefore, there are no elements of $\langle \gamma \rangle$ in $v_{1}$. Now, consider $w=\sum_{j=2}^{r}{\gamma^{n_{i_{j}}}\beta_{i_{j}}h_{i_{j}-1}^{k-1}}=(w_1,\dots,w_{L_\beta(k-1)})$. Since $\nu(c')>0$, we have that $\nu(w)>0$. We can permute the coordinates of $v_2$ so that we have $(\mathbf{z}_{(1)}+(w_1,\dots,w_1),\dots,$ $\mathbf{z}_{(1)}+(w_{L_\beta(k-1)},\dots,$ $w_{L_\beta(k-1)}))$. For $i\in\{1,\dots,$ $L_\beta(k-1)\}$,  $\mathbf{z}_{(1)}+$ $(w_i,\dots,w_i)$ is a permutation of $\mathbf{z}_{(1)}$ by Remark \ref{remark:sumpermv}, so each element in $\langle \gamma\rangle$ appears once. Therefore, in $v_2$, each element in $\langle \gamma\rangle$ appears $L_\beta(k-1)$ times.
\end{proof}


\begin{corollary} \label{coro:HamWeightCodewordBeta}
Let $c\in \mathcal{S}_{k}^{\beta} \setminus \{\zero\}$ with $k\geq 2$. Then,
$w_{H}(c)=L_{\beta}(k)-q^{\nu(c)}L_{\beta}(k-1).$
\end{corollary}

\begin{proof}
It follows from Proposition \ref{prop:2.10} and the fact that $0\in \langle q^{\nu(c)+1} \rangle$ for every valuation $\nu(c)\in\{0, 1, \dots, s-1\}$. 
\end{proof}

\begin{corollary} \label{coro:MinHamWeightSimplexBeta}
The minimum Hamming distance of $\mathcal{S}_{k}^{\beta}
$ is $d_{H}(\mathcal{S}_{k}^{\beta})=q^{s(k-1)}.$
\end{corollary}

Again, since the Hamming weight of any nonzero codeword of $\cS_k^\beta$ is determined by its valuation by Corollary \ref{coro:HamWeightCodewordBeta}, we can apply the same argument as for $\mathcal{S}_{k}^{\alpha}$, given in Theorem \ref{theo:HWESimplexAlpha}, to determine the Hamming weight distribution for the linear simplex $\beta$ code $\mathcal{S}_{k}^{\beta}$, and we obtain the following result.



\begin{theorem} \label{theo:HWESimplexBeta}
The Hamming weight enumerator for $\mathcal{S}_{k}^{\beta}$ with $k\geq 2$ is
\begin{align*}
W_{\mathcal{S}_{k}^{\beta}}(X,Y)=
&\sum_{j=0}^{s-1}A_{w_j}X^{L_{\beta}(k)-w_j}Y^{w_j}+X^{L_{\beta}(k)}, 
\end{align*}
where $w_j=L_{\beta}(k)-q^{j}L_{\beta}(k-1)$ and $A_{w_j}=q^{k(s-j)}-q^{k((s-j)-1)}$. The Hamming weight enumerator for $\cS_1^{\beta}$ is $W_{\cS_1^{\beta}}(X,Y)=(q^s-1)Y+X$.
\end{theorem}

Note that, while $\mathcal{S}_{k}^{\beta}$ offers less redundancy than $\mathcal{S}_{k}^{\alpha}$, since its length is smaller and both have the same number of codewords, it has smaller minimum distance as well, as $d_{H}(\mathcal{S}_{k}^{\beta})=q^{s(k-1)}\leq q^{sk-1}\leq (q-1)q^{sk-1}=d_{H}(\mathcal{S}_{k}^{\alpha})$. Moreover, we have that $d_{H}(\mathcal{S}_{k}^{\beta})<d_{H}(\mathcal{S}_{k}^{\alpha})$ unless $q=2$ and $s=1$.

Proposition \ref{prop:2.10} also allows us to calculate how many elements of each ideal of $R$ there are in any codeword of $\mathcal{S}_{k}^{\beta}$, which gives us the homogeneous weight of any codeword and the homogeneous weight distribution of the code as well.

\begin{proposition} \label{prop:2.15}
Let $c\in\mathcal{S}_{k}^{\beta}$ with $k\geq 1$. Then, we have that
$$w_{Hom}(c)=\begin{cases} q^{sk-1} & \text{if } \nu(c)=s-1, \\
  q^{sk-k-1}(q^{k}-1) & \text{if } \nu(c)<s-1, \\
  0 &\text{otherwise.}
  \end{cases}$$
  \end{proposition}
\begin{proof}
For $k=1$, the result follows from the definition of the homogeneous weight, as $\cS_{1}^{\beta}=R$.
Now, consider $k\geq 2$. We start by proving the case $\nu(c)=s-1$. By Proposition \ref{prop:2.10}, we have that $c$ has $q^{s-1}L_{\beta}(k-1)$ zeroes and $q^{s(k-1)}$ nonzero coordinates in $\langle \gamma^{s-1} \rangle$. Therefore, in this case, we have that $w_{Hom}(c)=q^{s(k-1)}q^{s-1}=q^{sk-1}$.

If $\nu(c)<s-1$, then we have that $s>1$, since $\nu(c)\geq 0$. This means that $\langle \gamma^{s-1} \rangle \subseteq \langle \gamma^{\nu(c)+1} \rangle$, and neither $\langle \gamma^{\nu(c)+1} \rangle$ nor $\langle \gamma^{s-1} \rangle $ are equal to $\{\zero \}$. By Proposition \ref{prop:2.10}, there are $q^{s(k-1)}$ coordinates of $c$ that belong to $\langle \gamma^{\nu(c)} \rangle \setminus \langle \gamma^{\nu(c)+1} \rangle$. Each of these coordinates has homogeneous weight $(q-1)q^{s-2}$, so their total homogeneous weight is  $(q-1)q^{s-2}q^{s(k-1)}$.


Again, by Proposition \ref{prop:2.10}, in the remaining coordinates, each element of $\langle \gamma^{\nu(c)+1} \rangle$  is repeated $q^{\nu(c)}L_{\beta}(k-1)$ times. Out of the total of $q^{s-1-\nu(c)}$ different elements in $\langle \gamma^{\nu(c)+1}\rangle$, the $q-1$ nonzero elements of $\langle \gamma^{s-1} \rangle$ have homogeneous weight $q^{s-1}$, and the others have weight $(q-1)q^{s-2}$. Therefore, we have that these elements add up to a total homogeneous weight of 
$\big((q-1)q^{s-1}+(q^{s-1-\nu(c)}-q)(q-1)q^{s-2}\big)q^{\nu(c)}L_{\beta}(k-1)=(q-1)q^{s-2}q^{s-1}L_{\beta}(k-1)$. 
Consequently, $w_{Hom}(c)=(q-1)q^{s-2}\big(q^{s(k-1)}+q^{s-1}L_{\beta}(k-1)\big)=(q-1)q^{s-2}L_{\beta}(k)$ by applying (\ref{eq:recsimp}). Finally, using that $L_{\beta}(k)=q^{(s-1)(k-1)}\frac{q^{k}-1}{q-1}$ by Proposition \ref{prop:3.2}, we obtain $w_{Hom}(c)=q^{s-2+sk-s-k+1}(q^{k}-1)=q^{sk-k-1}(q^{k}-1)$.

Finally, if $\nu(c)>s-1$, then $c=\zero$ and $w_{Hom}(c)=0$. 
\end{proof}

\begin{corollary} \label{cor:minhombeta}
The minimum homogeneous weight of $\mathcal{S}_{k}^{\beta}$ is $d_{Hom}(\mathcal{S}_{k}^{\beta})=q^{sk-k-1}(q^{k}-1).$
\end{corollary}


\medskip

Recall that \(\Phi_{R}:R\longrightarrow \F_q^{q^{s-1}}\) is an isometry between the homogeneous and Hamming weights \cite{Greferath}. Therefore, as in the case of $\cS_k^\alpha$, we can deduce the fundamental parameters and the Hamming weight distribution of the \(R\)-linear code \(\Phi_{R}(\mathcal{S}_{k}^{\beta})\) over \(\F_q\). Moreover, since \(R\)-linear codes are distance-invariant \cite{Greferath}, this also determines the Hamming distance distribution of \(\Phi_{R}(\mathcal{S}_{k}^{\beta})\).



\begin{theorem} \label{theo:simplexBetaParameters}
 The $R$-linear code $C=\Phi_{R}(\mathcal{S}_{k}^{\beta})$ with $k\geq 1$ is a $(q^{(s-1)k}\frac{q^{k}-1}{q-1}, q^{sk}, q^{sk-k-1}(q^{k}-1))$ code over $\mathbb{F}_q$. Moreover, the Hamming weight enumerator for $C$ is 
 $W_{C}(X,Y)=\sum_{\ell=0}^{W(C)}{A_{\ell}X^{N-\ell}Y^\ell}$, where $N=q^{(s-1)k}\frac{q^{k}-1}{q-1}$ and 
$$A_\ell=\begin{cases} 1 & \text{ if } \ell=0, \\
q^{k}-1 & \text{ if } \ell=q^{sk-1}, \\
q^{sk}-q^{k} &\text{ if } \ell=q^{sk-k-1}(q^{k}-1), \\
0 &\text{ otherwise}.
\end{cases}$$
\end{theorem}

\begin{proof}
The length of $C=\Phi_{R}(\mathcal{S}_{k}^{\beta})$ is 
$N=q^{s-1}L_{\beta}(k)=q^{(s-1)k}\frac{q^{k}-1}{q-1}$ by Proposition \ref{prop:3.2}. The number of codewords is $|\Phi_{R}(\mathcal{S}_{k}^{\beta})|=|\mathcal{S}_{k}^{\beta}|=q^{sk}$, and the minimum Hamming distance of $C$ equals $q^{sk-k-1}(q^{k}-1)$ by Corollary \ref{cor:minhombeta}, since the homogeneous weight of $c\in \cS_k^\beta$ coincides with the Hamming weight of $\Phi_R(c)\in C$.

Since the Hamming distance distribution of $C$ coincides with the homogeneous weight distribution of $\mathcal{S}_{k}^{\beta}$, we just need to count the number of codewords of $\cS_{k}^{\beta}$ that there are for every possible homogeneous weight.
Clearly, the only codeword of weight $0$ is the zero vector. The codewords of homogeneous weight $q^{sk-1}$ are those with valuation $s-1$ by Proposition \ref{prop:2.15}. By Lemma \ref{Lemma:valuationFree}, there are $q^{k(s-j)}-q^{k(s-j-1)}$ of them for $j=s-1$, so $q^{k}-1$. As for the codewords of homogeneous weight $q^{sk-k-1}(q^{k}-1)$, we know that there must be  $q^{sk}-1-(q^{k}-1)=q^{sk}-q^{k}$ of them.
\end{proof}

Table \ref{table:SimplexBeta} summarizes the Hamming and homogeneous weight distributions of $\cS_k^{\beta}$.


\begin{table}[h]

     \centering
   
     \begin{tabular}{|c|c|c|c|}
         \hline
          Valuation $\nu(c)$ &$w_{H}(c)$ &$w_{Hom}(c)$ &Number of codewords  \\
          \hline 
          $0\leq j\leq s-2$ &$L_{\beta}(k)-q^{j}L_{\beta}(k-1)$  &$q^{sk-k-1}(q^k-1)$ &$q^{k(s-j)}-q^{k((s-j)-1)}$\\
          \hline
          $s-1$ &$L_{\beta}(k)-q^{s-1}L_{\beta}(k-1)$ &$q^{sk-1}$ &$q^{k}-1$\\
          \hline
          $\infty$ &0 &0 &1\\
          \hline
     \end{tabular}
     \caption{Hamming and homogeneous weight distributions of $\cS_k^{\beta}$ in terms of the valuation of its codewords. Recall that $L_{\beta}(k)=q^{(s-1)(k-1)}\frac{q^k-1}{q-1}.$}
     \label{table:SimplexBeta}
\end{table}   

\vspace{-1truecm}


   


\begin{remark}
Consider the case $s=1$, so that $R=\F_q$. Recall that, in this case, we have that the homogeneous weight and the Hamming weight coincide and $\Phi_{R}$ is the identity map. Note that, by Theorem~\ref{theo:simplexBetaParameters}, $\cS_k^{\beta}$ is a code over $\F_q$ of length $\frac{q^k-1}{q-1}$ such that all nonzero codewords have Hamming weight $q^{k-1}$. Therefore, $\cS_k^{\beta}$ is a $[\frac{q^k-1}{q-1},k,q^{k-1}]$ $q$-ary linear code. In particular, the construction of its generator matrix detailed in Definition \ref{def:2.5} proves that $\cS_k^{\beta}$ is the classical $q$-ary simplex code.
\end{remark}

Finally, we present several detailed examples for small values of $q$, $s$, and $k$, illustrating the theoretical results for $\cS_k^{\beta}$ and providing the corresponding Hamming and homogeneous weight distributions.

\begin{example}
Let $R=\Z_{4}$ and $k=2$. In this case, $q=2=s$, and the homogeneous weight coincides with the Lee weight.   By Definition \ref{def:2.5}, the generator matrix of the linear simplex $\beta$ code $\cS_2^{\beta}$ over $\Z_4$ is
$$G_{2}^{\beta}=\left(\begin{array}{ccc}
\mathbf{1} &0 &2\\
0\,1\, 2\, 3       &1 &1
\end{array}
\right)=\left(\begin{array}{c}
h_1^{2}\\ 
h_2^{2}\end{array}\right).$$
Note that the length of $\cS_2^{\beta}$ is equal to $L_{\beta}(2)=4+2=6$, as shown in Proposition \ref{prop:3.2}. Therefore, $\cS_2^{\beta}$ has type $(6;2,0)$, as shown in Proposition \ref{prop:3.3}.

The Hamming weight of a nonzero codeword $c$ is $w_0=6-1=5$ or $w_1=6-2^{1}=4$, depending on whether $c$ has valuation $0$ or $1$, respectively. For example, $w_H(h_1^2)=w_H((111102))=5$ and $w_{H}(2h_2^2)=w_H((020222))=4$. Since $A_{w_{0}}=12$ and $A_{w_1}=3$, the Hamming weight enumerator of $\cS_2^{\beta}$ is $W_{\cS_2^{\beta}}(X,Y)=12X^1Y^5+3X^{2}Y^{4}+X^{6}$, as shown in Theorem \ref{theo:HWESimplexBeta}.

The homogeneous weight of a nonzero codeword $c$  is $w_{0}=6$ or $w_1=8$, depending on whether $\nu(c)=0$ or $\nu(c)=1$, respectively.  Recall that the elements $1$ and $3$ of $\Z_4$ have homogeneous weight $1$, whereas $2$ has homogeneous weight $2$. For example, $w_{Hom}(h_1^2)=4+2=6$, and $w_{Hom}(2h_2^2)=4\cdot 2=8$.
Since $A_{w_0}=12$ and $A_{w_1}=3$, and the homogeneous weight enumerator of $\cS_2^{\beta}$ coincides with the Hamming weight enumerator of its Gray map image $C=\Phi_R(\cS_2^{\beta})$, we obtain $W_C(X,Y)=12X^{6}Y^{6}+3X^{4}Y^{8}+X^{12}$. Consequently, $C$ is a binary $(12,16,6)$ code, as shown in Theorem \ref{theo:simplexBetaParameters}.
\end{example}

\begin{example}
Let $R=GR(4,2)$, as in Example \ref{ex:SimplexAlphaGR42}, and $k=1$. In this case, $q=4$ and $s=2$. The generator matrix of $\cS_1^{\beta}$  is $G_1^{\beta}=(1)$. Note that $L_{\beta}(1)=1$ and $\cS_1^{\beta}=R$. Therefore,  the code $\cS_1^{\beta}$ has type $(1;1,0)$, as stated in Proposition \ref{prop:3.3}.

The Hamming weight of every nonzero codeword is $1$.  Therefore, $W_{\cS_1^{\beta}}(X,Y)=15Y+X$, as shown in Theorem \ref{theo:HWESimplexBeta}.
The homogeneous weight of a nonzero codeword $c$ is $w_0=4^{0}\cdot(4-1)=3$ or $w_1=4^1=4$, depending on whether $\nu(c)=0$ or $\nu(c)=1$, respectively. For example, $w_{Hom}(\omega)=3$, and $w_{Hom}(2+2\omega)=4$. 
Consequently, $C=\Phi_{R}(\cS_1^{\beta})$ is a $(4,16,3)$ code over $\F_4$ with weight enumerator $W_C(X,Y)=12XY^{3}+3Y^{4}+X^{4}$ as shown in Theorem \ref{theo:simplexBetaParameters}. Indeed,  $C=\Phi_R(R)$ is the first-order Reed-Muller code over $\F_4$, as mentioned after the definition of $\Phi_R$ in (\ref{eq:Greferath}).
\end{example}



To conclude this section, we check whether the linear simplex $\alpha$ and $\beta$ codes, $\mathcal{S}_{k}^{\alpha}$ and $\mathcal{S}_{k}^{\beta}$, are optimal with respect to the Griesmer bound. This bound, stated  in \cite{ShiromotoG} for linear codes over finite Frobenius local rings, applies in particular to linear codes over finite chain rings, since every finite chain ring is Frobenius. A code meeting this bound is called {\it optimal} with respect to the Griesmer bound.

\begin{proposition}[Griesmer Bound]{\cite[Theorem 2.2]{ShiromotoG}}
Let $R$ be a finite Frobenius local ring $R$ with residue field $\mathbb{F}_{q}$, and let $C_R$ be a free linear code  over $R$ of length $n$, generated by $k$ linearly independent codewords. Then,
\begin{equation} \label{eq:GriesmerBound}
n\geq \sum_{i=0}^{k-1}{\lceil \frac{d_{H}(C_R)} {q^{i}}}\rceil.
\end{equation}
\end{proposition}

Let $C_R$ be a linear code over $R$ of length $n$. We define its canonical projection onto the residue field $\mathbb{F}_q$ by $\overline{C_R}=\{\overline{c}\mid c\in C_R\}$, where $\overline{c}$ denotes the componentwise canonical projection of $c\in R^n$ onto $\mathbb{F}_q^n$.  Then $\overline{C_R}$ is a linear code of length $n$ over $\mathbb{F}_q$. Moreover, if $G$ is a generator matrix of $C_R$, then $\overline{G}$ generates $\overline{C_R}$.
It was proved in \cite[Theorem 4.2]{SalageanaHamming} that, when $\overline{C_R}\neq \{\zero\}$, $d_H(\overline{C_R})\leq d_H(C_R).$
Furthermore, if $C_R$ is free, then $d_H(\overline{C_R})=d_H(C_R).$
Thus, $C_R$ and $\overline{C_R}$ have the same length and, whenever $C_R$ is free, the same dimension and minimum Hamming distance. Consequently, they yield the same Griesmer bound. 

In our setting, both $\overline{\mathcal{S}_k^{\alpha}}$ and $\overline{\mathcal{S}_k^{\beta}}$ are nonzero codes, since the generator matrices $G_k^{\alpha}$ and $G_k^{\beta}$ contain columns with at least one unit entry. Moreover, both $\mathcal{S}_k^{\alpha}$ and $\mathcal{S}_k^{\beta}$ are free codes.  Therefore, we can study the optimality of these codes with respect to the Griesmer bound by considering their projections over $\mathbb{F}_q$. We first show that $\mathcal{S}_k^{\alpha}$ is not optimal by proving that the equality in (\ref{eq:GriesmerBound}) is not attained.

\begin{remark} \label{rmk:simplexalphagriesmer}
The linear simplex code $\mathcal{S}_{k}^{\alpha}$ satisfies that  $d_{H}(\mathcal{S}_{k}^{\alpha})=(q-1)q^{sk-1}$ by Corollary \ref{coro:MinHamWeightSimplexAlpha}. Therefore, 
\begin{equation} \begin{split} 
&\sum_{i=0}^{k-1}{\lceil\frac{d_{H}(\mathcal{S}_{k}^{\alpha})}{q^{i}}}\rceil=\sum_{i=0}^{k-1}{\frac{(q-1)q^{sk-1}}{q^{i}}}=(q-1)\sum_{i=0}^{k-1}{q^{sk-1-i}}
=(q-1)\sum_{i=0}^{k-1}{q^{(s-1)k+i}}\\ &=(q-1)q^{(s-1)k}\sum_{i=0}^{k-1}{q^{i}}=(q-1)q^{(s-1)k}\frac{q^{k}-1}{q-1}=q^{sk}-q^{(s-1)k}.
\end{split} \end{equation}
By Proposition \ref{prop:3.1}, the length of $\mathcal{S}_{k}^{\alpha}$ is $q^{sk}$, so these codes do not meet the Griesmer bound (\ref{eq:GriesmerBound}) since  $q^{sk}-q^{(s-1)k}<q^{sk}$.
\end{remark}

We now explain this lack of optimality. Since the columns of $G_k^{\alpha}$ contain every element of $R^k$, they contain every element of $\langle\gamma\rangle^k$. The canonical projection of each column in $\langle\gamma\rangle^k$ is the zero vector, resulting in zero columns that increase the length of  $\overline{\mathcal{S}_k^{\alpha}}$ without affecting its minimum Hamming distance. Thus, the original code is longer than necessary to attain the Griesmer bound.


We now show that the code \(\mathcal{S}'_k\), obtained from \(G_k^\alpha\) by removing the columns in \(\langle\gamma\rangle^k\), is optimal with respect to the Griesmer bound. By Remark~\ref{rmk:idealcount}, its length is \(q^{sk}-q^{(s-1)k}\), and $\mathcal{S}'_k$ is free over $R$, by the same argument as in Proposition~\ref{prop:3.1}. Therefore, both \(\mathcal{S}'_k\) and \(\mathcal{S}_k^\alpha\) are free over \(R\), so they have the same dimension, and their canonical projections have the same minimum Hamming distance. Hence,
yield the same right-hand side of the Griesmer bound, as computed in Remark~\ref{rmk:simplexalphagriesmer}. Since \(\mathcal{S}'_k\) attains this bound, it is optimal.



Finally, we consider the code $\cS_k^{\beta}$. By construction, every column of $G_k^{\beta}$ contains the entry $1$. Hence, $\overline{G}_k^{\beta}$ has no zero columns, and the canonical projection does not introduce redundant coordinates that can be removed without affecting the minimum Hamming distance. We next show that $\mathcal{S}_k^{\beta}$ attains the Griesmer bound, so it is optimal.

\begin{proposition} \label{prop:griesmerbeta}
The linear simplex code $\mathcal{S}_{k}^{\beta}$ over $R$ is optimal with respect to the Griesmer bound for any finite chain ring $R$ and $k\geq 1$.
\end{proposition}
\begin{proof}
By Corollary \ref{coro:MinHamWeightSimplexBeta}, $d_{H}(\mathcal{S}_{k}^{\beta})=q^{s(k-1)}$. Therefore, 
\begin{equation} 
\sum_{i=0}^{k-1}{\lceil\frac{d_{H}(\mathcal{S}_{k}^{\beta})}{q^{i}}}\rceil=\sum_{i=0}^{k-1}{q^{s(k-1)-i}}=\sum_{i=0}^{k-1}{q^{(s-1)(k-1)+i}}
=q^{(s-1)(k-1)}\frac{q^{k}-1}{q-1}=L_{\beta}(k)
\end{equation}
by Proposition \ref{prop:3.2}, and these codes meet the Griesmer bound (\ref{eq:GriesmerBound}).
\end{proof}

\section{On $\Z_{p^s}$-additive simplex codes}
\label{sec:SimplexZps}


In this section, we list the results obtained in our work for the case $R=\Z_{p^{s}}$. Among linear codes over finite chain rings, one of the most extensively studied families is that of codes over the ring $\Z_{p^s}$ \cite{Dipak23,Dipak22,Shi,Sloane}. In the case of $\mathbb{Z}_{p^s}$-additive codes, results can also be expressed in terms of the orders of the codewords, providing a natural framework for describing their structural properties. Moreover, as  for codes over $\mathbb{Z}_{2^s}$ \cite{Nonlinearity}, the generator matrices of linear simplex codes over $\mathbb{Z}_{p^s}$ are related to the construction of generator matrices of $\mathbb{Z}_{p^s}$-additive GH codes. Presenting these results in the setting of $\mathbb{Z}_{p^s}$ makes the generalization from $\mathbb{Z}_{2^s}$ more transparent and allows us to present the general results of this work in a more familiar setting.

When $R=\Z_{p^{s}}$, we have $\gamma=p$, so $K=\Z_{p}$ and $q=p$. 
Moreover, since the structure of a $\Z_{p^{s}}$-module is  completely determined by its underlying abelian group structure, any property  expressed in terms of the valuation of a vector $x\in\Z_{p^{s}}^{n}$ can be reformulated in terms of its order, and vice versa, since $\nu(x)=s-\log_{p}(o(x))$, or equivalently, $o(x)=p^{s-\nu(x)}$. Recall that, for $x\in R^{n}$, the {\it order} of $x$, denoted by $o(x)$, is the minimum natural number $m>0$ such that $mx=\zero$.

We show that, over the ring  $\Z_{p^{s}}$, the linear simplex codes $\cS_{k}^\alpha$ and $\cS_{k}^\beta$ are closely related to $\Z_{p^{s}}$-additive GH codes of type $(n;k+1,0,\dots,0)$. Recall that a $\Z_{p^s}$-additive GH code $\cH$ is a linear code over $\Z_{p^s}$ such that its corresponding $\Z_{p^s}$-linear code $\Phi_R(\cH)$ is a GH code. The Gray map $\Phi_{R}$ for $\Z_{p^{s}}$ is the generalized Carlet Gray map \cite{Carlet2k} used in works such as \cite{Dipak22,Dipak23,Hadamard2,HadamardZ8,Nonlinearity,AdrianZp}.

For any sequence $t_{1},\dots,t_{s}$ of nonnegative integers such that $t_{1}\geq1$, we can construct a $\Z_{p^{s}}$-additive GH code $\mathcal{H}^{t_{1},\dots,t_{s}}$ of type $(n;t_{1},\dots,t_{s})$, by constructing a generator matrix,  denoted by $A^{t_{1},\dots,t_{s}}$ \cite{Dipak22}.  We focus on matrices of the form $A^{t_1,0,\dots,0}$, which are generator matrices of the free $\Z_{p^{s}}$-additive GH codes $\cH^{t_1,0,\dots,0}$. The recursive construction is as follows. First, $A^{1,0,\dots,0}=(1)$, and for $t_1\geq 2$, we define
$$A^{t_1,0,\dots,0}=\left(\begin{array}{ccccc}
 \zero &\mathbf{1} &\mathbf{2} &\dots &\mathbf{p^s-1}\\
 A &A &A &\dots &A
\end{array}\right),$$
where $A=A^{t_1-1,0,\dots,0}$. Note that the last row of $A^{t_1,0,\dots,0}$ has all entries equal to $1$.

By Definition \ref{def:simplexalpha}, for $R=\Z_{p^{s}}$, the code $\mathcal{S}_{k}^{\alpha}$ is defined as the linear code with generator matrix $G_{k}^{\alpha}$, where
$G_{1}^{\alpha}=\left(\begin{array}{ccccc} 0 &1 &2 &\dots &p^{s}-1\end{array}\right)$ and, for $k>1$, $$G_{k}^{\alpha}=\left(\begin{array}{ccccc}
\mathbf{0} &\mathbf{1} &\mathbf{2} & \dots & \mathbf{p^{s}-1}\\
G_{k-1}^{\alpha} & G_{k-1}^{\alpha} & G_{k-1}^{\alpha} & \dots & G_{k-1}^{\alpha}
\end{array}\right).$$ 
Now, consider $\tilde{A}^{k+1,0,\dots,0}$, which is obtained from the matrix $A^{k+1,0,\dots,0}$ by removing the all-one row. Comparing the constructions of $G_k^{\alpha}$ and $A^{k+1,0,\dots,0}$, we see that $G^\alpha_k = \tilde{A}^{k+1,0,\dots,0}$.
By Definition \ref{def:2.5}, the code $\mathcal{S}_{k}^{\beta}$ is the linear code generated by the matrix $G_{k}^{\beta}$, where
$G_{1}^{\beta}=(1)$ and, for $k>1$,
$$G_{k}^{\beta}=\left(\begin{array}{cccccc}
\mathbf{1} &\mathbf{0} &1\mathbf{p} &2\mathbf{p} & \dots &(p^{s-1}-1)\mathbf{p}\\
G_{k-1}^{\alpha} & G_{k-1}^{\beta} & G_{k-1}^{\beta} & G_{k-1}^{\beta} &\dots &G_{k-1}^{\beta} 
\end{array}\right).$$ 
In Examples \ref{ex:SimplexAlphaZ9} and \ref{ex:SimplexBetaZ9}, the generator matrices $G_2^\alpha$ and $G_2^\beta$ over $\Z_9$ are shown. 

Next, we give the Hamming and homogeneous weight distributions of the linear simplex codes $\mathcal{S}_{k}^{\alpha}$ and $\mathcal{S}_{k}^{\beta}$ over $\Z_{p^s}$ in terms of the order of the codewords. As corollaries of Propositions \ref{prop:2.7} and \ref{prop:2.10}, we obtain Corollaries \ref{coro:ZpsHammingAlpha} and \ref{coro:ZpsHammingBeta}, which generalize the results given by Gupta for linear simplex codes over $\Z_{2^{s}}$ in \cite{Gupta}.
  
\begin{corollary}\label{coro:ZpsHammingAlpha}
Let $c\in \mathcal{S}_{k}^{\alpha}\setminus \{\zero\}$. Then, $w_{H}(c)=p^{sk}-\frac{p^s}{ord(c)}p^{s(k-1)}.$    
\end{corollary}

\begin{corollary}  \label{coro:ZpsHammingBeta}
Let $c\in \mathcal{S}_{k}^{\beta} \setminus \{\zero\}$ with $k\geq 2$. Then,
$w_{H}(c)=L_{\beta}(k)-\frac{p^{s}}{ord(c)}L_{\beta}(k-1).$
\end{corollary}

The minimum Hamming distance of $\mathcal{S}_{k}^{\alpha}$ is $(p-1)p^{sk-1}$, which equals $\varphi(p^{sk})$, where $\varphi$ is Euler's totient function. By expressing  Theorem \ref{theo:HWESimplexAlpha} in terms of the order, we have that the number of codewords of order $p^i$, $i\in \{1,\dots,s\}$, that is, of Hamming weight $\bar w_i=p^{sk}-p^{sk-i}$ is $A_{\bar w_i}=p^{ki}-p^{k(i-1)}$. Since the homogeneous weight of a codeword of $\mathcal{S}_{k}^{\alpha}$ does not depend on its order,  Theorem \ref{theo:simplexAlphaParameters}  directly yields the following result, giving the fundamental parameters and Hamming weight distribution of its Gray map image $\Phi_R(\cS_k^\alpha)$.

\begin{corollary}
\label{prop:simplexAlphaParametersZps}
The $\Z_{p^s}$-linear code $\Phi_{R}(\mathcal{S}_{k}^\alpha)$ is a $(p^{s(k+1)-1}, p^{sk}, p^{s(k+1)-2}(p-1))$ code over $\Z_{p}$,
having all codewords of Hamming weight $p^{s(k+1)-2}(p-1)$, except the all-zero codeword.
\end{corollary}

The minimum Hamming distance of $\mathcal{S}_{k}^{\beta}$ is $p^{s(k-1)}$. By Theorem \ref{theo:HWESimplexBeta}, the number of codewords of order $p^i$, $i\in \{1,\dots,s\}$, that is of Hamming weight $\bar w_i=L_\beta(k)-p^{s-i}L_\beta(k-1)$ is
$A_{\bar w_i}=p^{ki}-p^{k(i-1)}$. Furthermore, by expressing  the results given in Proposition \ref{prop:2.15} and Theorem  \ref{theo:simplexBetaParameters} for the homogeneous weight of any codeword of $\mathcal{S}_{k}^{\beta}$ in terms of the order, we can also obtain the fundamental parameters and Hamming weight distribution of $\Phi_{R}(\mathcal{S}_{k}^{\beta})$.


\begin{corollary} 
Let $c\in\mathcal{S}_{k}^{\beta}$ with $k\geq 2$. Then, we have that
$$w_{Hom}(c)=\begin{cases} p^{sk-1} & \text{if } ord(c)=p, \\
  p^{sk-k-1}(p^{k}-1) & \text{if } ord(c)>p, \\
  0 &\text{otherwise.}
  \end{cases}$$
\end{corollary}

\begin{corollary} 
 The $\Z_{p^s}$-linear code $\Phi_{R}(\mathcal{S}_{k}^{\beta})$ with $k\geq 2$ is a $(p^{(s-1)k}\frac{p^{k}-1}{p-1},$ $p^{sk},$  $p^{sk-k-1}(p^{k}-1))$ code over $\Z_p$,
having $p^{k}-1$ codewords of Hamming weight $p^{sk-1}$, $p^{sk}-p^{k}$ codewords of Hamming weight $p^{sk-k-1}(p^{k}-1)$ and one of Hamming weight $0$.
\end{corollary}

Note that, for $p=2$, the results on the Hamming and homogeneous weight distributions of $\cS_k^{\alpha}$ given in Propositions 3.3 and 3.4 and Theorems 3.1 and 3.2, as well as those on the Hamming and homogeneous weight distributions of $\cS_k^{\beta}$ given in Propositions 3.5 and 3.6 and Theorems 3.3 and 3.4 coincide with those obtained in \cite{Gupta} for $\mathbb{Z}_{2^s}$-additive simplex codes of types $\alpha$ and $\beta$, respectively. Furthermore, Proposition~3.7 generalizes the Griesmer bound result established in \cite{Gupta} for $\mathbb{Z}_{2^s}$-additive simplex codes of type $\beta$.

\section{Conclusions and further research}
\label{sec:conclusion}

In this work, we have introduced two families of simplex codes over a finite chain ring $R$, $\cS_k^{\alpha}$ and $\cS_k^{\beta}$, generalizing the simplex codes of types $\alpha$ and $\beta$ introduced in \cite{Gupta} for $R=\Z_{2^s}$. We have calculated their lengths and cardinality, showing that they are free modules. Moreover, we have derived explicit formulas for their Hamming and homogeneous weight distributions in terms of the parameter $k$ and the ring parameters $q$ and $s$, as well as the valuations of the codewords. Since the generalized Gray map $\Phi_{R}$ is an isometry between the metric spaces $(R^{n},d_{Hom})$ and $(\F_{q}^{q^{s-1}n},d_{H})$, and its images are distance-invariant codes, the homogeneous weight distributions of $\cS_k^{\alpha}$ and $\cS_{k}^{\beta}$ also allow us to determine the Hamming distance distributions of their Gray map images, $\Phi_R(\cS_k^{\alpha})$ and $\Phi_{R}(\cS_k^{\beta})$, respectively. Finally, we have studied whether these codes attain the Griesmer bound for codes over rings given in \cite{ShiromotoG}.


It is worth mentioning that, given the special properties of finite left and right chain rings \cite{NechPIR}, the ideas developed in this paper can be extended to the noncommutative setting. In particular, by defining the left (right) linear codes $\cS_k^{\alpha}$ and $\cS_k^{\beta}$ over a left (right) finite chain ring $R$ using analogous generator matrices, we can define a valuation function on $R$ and apply the same techniques to derive their Hamming and homogeneous weight distributions.

A possible direction for future research is to  determine the fundamental parameters and derive explicit formulas for the Hamming and homogeneous weight distributions of the linear MacDonald codes $\cM_{k,u}^{\alpha}$ and $\cM_{k,u}^{\beta}$ over any finite chain ring $R$. These codes, introduced in \cite{Nonlinearity} for $R=\Z_{2^s}$, are obtained by puncturing the corresponding linear simplex codes $\cS_k^\alpha$ and $\cS_k^\beta$. 
Another potential research avenue is the  construction of $R$-linear GH codes that generalize the construction given in \cite{Dipak23} for $R=\Z_{p^{s}}$, while also exploring their connection with linear simplex codes over $R$.

\section*{Acknowledgments}






The material in this paper was presented in part at V Jornadas de la red MatSI (https://web.ua.es/es/matsi2024/) held from 5-7 June 2024 at the University of Alicante, Spain. This work has been partially supported by the Spanish Government under grants PID2022-137924NB-I00 and PID2025-170476NB-I00.



The authors declare there are no conflicts of interest.

\bibliography{Bibliography}

\end{document}